\documentclass[lettersize,journal]{IEEEtran}
\usepackage{amsmath,amsfonts}
\usepackage{algorithmic}
\usepackage{algorithm}
\usepackage{array}
\usepackage[caption=false,font=normalsize,labelfont=sf,textfont=sf]{subfig}
\usepackage{textcomp}
\usepackage{stfloats}
\usepackage{url}
\usepackage{verbatim}
\usepackage{graphicx}
\usepackage{cite}
\usepackage{listings}
\usepackage{amsmath,amsfonts}
\usepackage{graphicx}
\usepackage{amsfonts}
\usepackage{array}
\usepackage{amssymb}
\usepackage{helvet}
\usepackage{color}
\usepackage{bm}
\usepackage{float}
\usepackage{cite}
\usepackage{multirow}
\usepackage{slashbox}
\usepackage{amssymb}
\usepackage[caption=false,font=normalsize,labelfont=sf,textfont=sf]{subfig}
\usepackage{amsmath,amsfonts}
\usepackage{algorithmic}
\usepackage{algorithm}
\usepackage{textcomp}
\usepackage{stfloats}
\usepackage{url}
\usepackage{verbatim}
\usepackage{graphicx}
\usepackage{cite}
\usepackage{amssymb}
\usepackage{makecell}

\newcommand{\RNum}[1]{\uppercase\expandafter{\romannumeral #1\relax}}

\newcounter{MYtempeqncnt}

\begin{document}

\title{Integrated Sensing and Communications  in \\ Clutter Environment}
\author{Hongliang Luo,  Yucong Wang, Dongqi Luo, Jianwei Zhao,  Huihui Wu,  Shaodan Ma and Feifei Gao
\thanks{Hongliang Luo,  Yucong Wang, Dongqi Luo, Huihui Wu and Feifei Gao are with 
Department of Automation, Tsinghua University, 
Beijing 100084, China (email: luohl23@mails.tsinghua.edu.cn;  wangyuco21@mails.tsinghua.edu.cn;
dongqiluo@gmail.com; hhwu1994@
mail.tsinghua.edu.cn; feifeigao@ieee.org).}
\thanks{
Jianwei Zhao is with the High-Tech Institute of Xi'an, Xi'an 710025, China~(e-mail: zhaojianeiep@163.com).}
\thanks{Shaodan Ma is with the State Key Laboratory of Internet of Things for Smart City and the Department of Electrical and Computer Engineering, University of Macau, Macao S.A.R. 999078, China (e-mail: shaodanma@um.edu.mo).
}
}



\maketitle

\begin{abstract}
In this paper, we propose a practical integrated sensing and communications (ISAC) framework to sense    dynamic targets from   clutter environment while ensuring users communications quality.
To implement  communications function and sensing function simultaneously, we  design multiple communications beams that can communicate with the users as well as one  sensing beam that can rotate and scan    the entire  space.
To minimize the interference of sensing beam on existing communications systems, we divide the service area into \emph{sensing beam for sensing (S4S) sector} and  \emph{communications beam for sensing (C4S) sector}, and provide beamforming design and power allocation optimization strategies for each type sector.
Unlike most existing ISAC studies that ignore the  interference of static environmental clutter on target sensing,
we construct a mixed sensing channel model that includes both static environment and dynamic targets.
When base station receives the echo signals,
it first 
filters out the interference from static environmental clutter and extracts the effective dynamic target echoes.
Then a complete and practical dynamic target sensing scheme is designed   to detect the presence of dynamic targets and to estimate their angles, distances, and velocities.
In particular,
dynamic target detection and angle estimation are realized through
 angle-Doppler spectrum estimation (ADSE) and joint detection  over multiple subcarriers  (MSJD), while distance and velocity estimation are realized through the extended subspace algorithm.
Simulation results  demonstrate the effectiveness of the proposed scheme and its superiority over the existing methods that ignore   environmental
 clutter.
\end{abstract}


\section{Introduction}
Integrated sensing and communications (ISAC) has recently attracted  significant  research interest  in the field of wireless communications\cite{202310141,9040264,ISAC1}.  The idea is to utilize    communication signals to sense various information in  real physical world, such as  architectural composition,  human activity, etc,
while ensuring  users communications quality.
Since ISAC  allows sensing systems and communication systems to share
the same hardware and spectrum resources, and can serve various  intelligent applications, such as vehicle-to-everything, digital twins,  etc., it has been deemed as one of the key technologies for sixth generation  (6G)   communications\cite{a11221,9755276}.

The ultimate functionality  of sensing is to construct the mapping relationship from  \emph{real physical world} to  \emph{digital twin world},
where the former includes 
 static environment (such as roads and buildings) and dynamic targets (such as pedestrians and vehicles).
Therefore,  sensing tasks in ISAC systems include  \emph{static environment sensing (SES)}  and \emph{dynamic targets sensing (DTS)} as shown in  left half of Fig.~1.
 The  changes of  static environment are usually slow, and thus
one can apply various environmental reconstruction  techniques to sense  static environment\cite{9727176,9516898,10182348}.
 However, since the  changes of dynamic targets are rapid, 
 it is necessary  to   detect the presence of dynamic targets and update their parameter estimates in real-time.
Generally, one can divide  DTS   problem into two categories:
1) sensing  cooperative communications users, e.g., mobile phones;  and
2) sensing  non-cooperative dynamic targets that are not communicating with   base station (BS), e.g., moving objects or mobile  users not in communication status.  

\begin{figure}[!t]
\centering
\includegraphics[width=90mm]{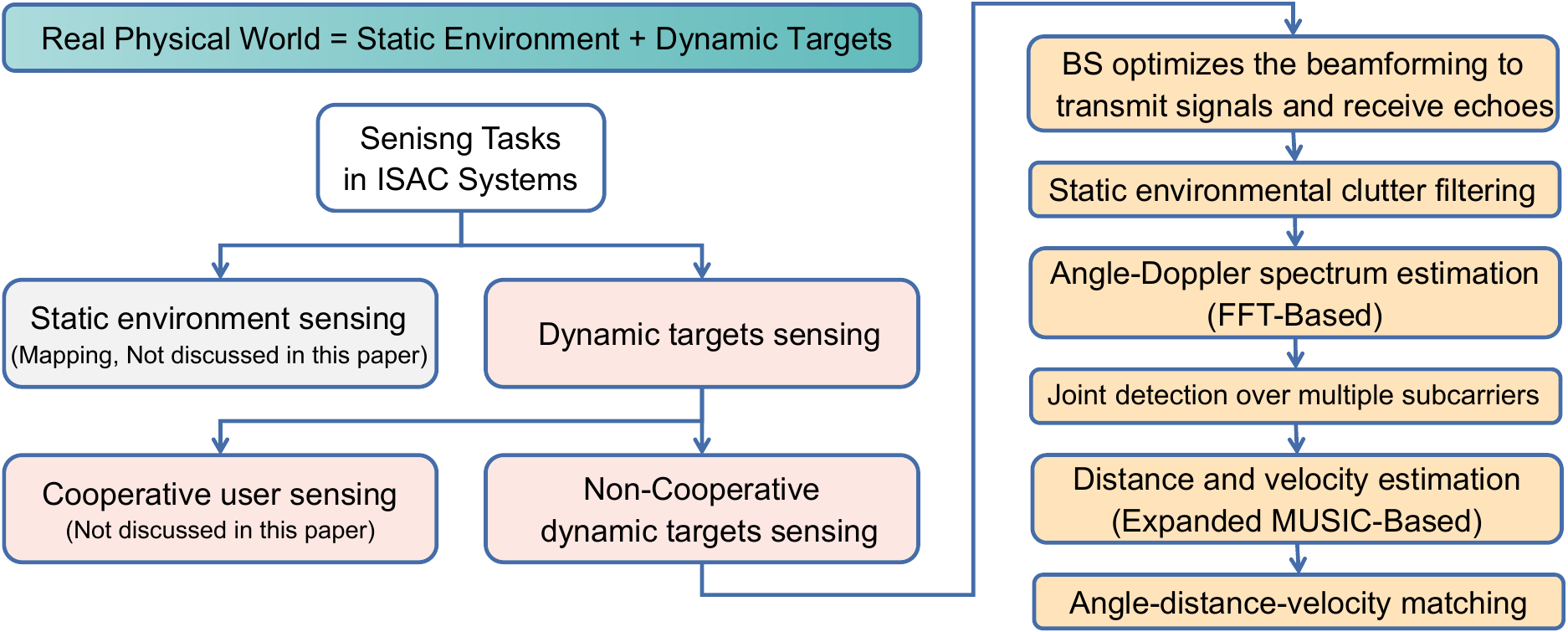}
\caption{ISAC sensing task classification and flowchart of the proposed sensing echo signals processing process.}
\label{fig_1}
\end{figure}

In addition, with the increasing demand for sensing, researchers are committed to making  ISAC systems possess ultra-high sensing accuracy.
Encouragingly,
 sensing in mmWave or Terahertz band with massive multiple input and  multiple output (MIMO) array can achieve higher accuracy thanks to its high directivity and high temporal resolution\cite{9782674,9013639}.
Therefore, the massive MIMO array based ISAC systems have been widely studied.
For the aspect of sensing  cooperative  users,  the users' positions can be estimated by processing the communications signals and  extracting the channel parameters, such as  time of arrival (TOA),      angle of departure (AOD), etc\cite{g333,g111,g222,9506874}.
On the other side, sensing  non-cooperative target has also been studied
 under  ISAC framework.
For example, 
Z.~Gao~\emph{et.~al.}  proposed an ISAC  system relying on  MIMO array,
which applied  compressed sampling   to facilitate
target sensing and other ISAC processing\cite{9898900}.
Z. Wang~\emph{et. al.} proposed a simultaneously transmitting and reflecting surface  enabled ISAC framework, in which the two-dimensional maximum likelihood estimation was utilized to estimate target angle\cite{10050406}.  
 Y. Li \emph{et. al.} proposed a two-stage algorithm  to estimate the positions and velocities of multiple targets with  orthogonal frequency division multiplexing (OFDM) signals\cite{8918315}. 
P.~Kumari \emph{et.~al.} proposed  an ISAC system working  for the internet of vehicles, which realized vehicle to vehicle  communication and full duplex radar sensing\cite{8114253}. X.~Chen~\emph{et.~al.}  proposed a multiple signal classification  based ISAC system that can  attain high estimation accuracy for dynamic target sensing\cite{10048770}.
However,  all these works ignore  the  interference caused by static environmental   clutter on dynamic target sensing, 
 which does not match  the real scenarios
where  the  targets are always
 submerged by the  interference of static environment
 and are difficult to separate from it,  thus limiting the sensing performance in the real scenarios and practical ISAC systems.

Meanwhile, for  ISAC systems based on MIMO arrays, 
one should  design the beamforming with multiple beams 
 to implement  communications function and sensing function simultaneously\cite{9124713,9854898,9850347,9761984}.
For example, 
C.~B.~Barneto \emph{et.~al.}   formulated one  beamforming design problem in  multi-user ISAC system,
aiming to maximize the  power at the sensing direction while remain certain  power at the communications directions\cite{99330000}.
H.~Luo \emph{et.~al.} proposed an ISAC system that combined BS and  reconfigurable intelligence surface (RIS), and  optimized the beamfomring matrix jointly to maximize the achievable sum-rate of communication users while meeting the sensing beam constraints\cite{9852716}.
X.~Wang \emph{et.~al.}  investigated the partially-connected  beamforming design for multi-user ISAC systems, aiming to \textcolor{black}{minimize} the 
Cramer-Rao  bound  of  DOA estimation
while ensuring communications performance\cite{9868348}.
However,  all 
these works require complex optimization processes and are difficult to implement  in practical systems.

\textcolor{black}{Based on the above analysis, the  existing researches on ISAC systems require complex optimization solutions in beamforming design and ignore the serious interference of environmental clutter in echo signal processing, making it difficult to deploy and apply in actual BS systems.}
In this paper, we propose a practical ISAC framework to sense the    dynamic targets from the  clutter environment while ensuring users communications quality.
The contributions of this paper are summarized as follows.

\begin{itemize}

\item We propose a practical ISAC framework to implement  communications  and sensing  simultaneously, in which 
 we  design multiple communications beams that can communicate with the users while assign  one  sensing beam that can rotate and scan the entire  space.
The entire process is summarized in the right half of Fig.~1.

\item
 BS should optimize the transmitting  beamforming during each time slot to transmit  signals.
Specifically,  to minimize the interference of sensing beam on existing communications systems, we divide the service area of  BS into \emph{sensing beam for sensing (S4S) sectors}   and \emph{communications beam for sensing (C4S) sectors}.
To avoid complex optimizations  that are difficult to implement in practical systems, we transform the precoding optimization problem into a power allocation problem for communications and sensing beams,
and provide effective power allocation  strategies for each type sector.

\item We construct  clutter environment model for  practical ISAC systems, and 
 then the echoes received by  BS includes both the echoes caused by  static environment and the echoes caused by  dynamic targets. 
To address the negative interference of static environmental clutter on dynamic target sensing, 
we first 
 filter out the interference from static environmental  clutter and  extract the effective dynamic target echoes. 
Then   angle-Doppler spectrum estimation (ADSE) and 
joint detection  over multiple subcarriers  (MSJD)
 are proposed to detect   dynamic targets  and to estimate their angles.
The extended subspace algorithm is utilized to estimate  distances and velocities of  dynamic targets.
Besides,  in order to obtain multiple parameters for each target, we design an angle-distance-velocity matching method to realize parameters matching.

\item 
Various simulations are provided to demonstrate the effectiveness of the proposed schemes and its superiority over  existing methods that ignore  
environmental clutter.

\end{itemize}

The remainder of this paper is organized as follows.
In Section \RNum{2}, we propose a practical ISAC framework that effectively integrates the sensing functions into the existing communications systems.
In Section~\RNum{3}, we provide a low complexity and practical beamforming design and power allocation strategy.
Then we design a complete and practical scheme 
to sense the   dynamic targets from the  clutter environment in Section \RNum{4}.
Simulation results and conclusions are given in Section~\RNum{5} and Section~\RNum{6},
respectively.

\begin{figure}[!t]
	\centering
	\includegraphics[width=77mm]{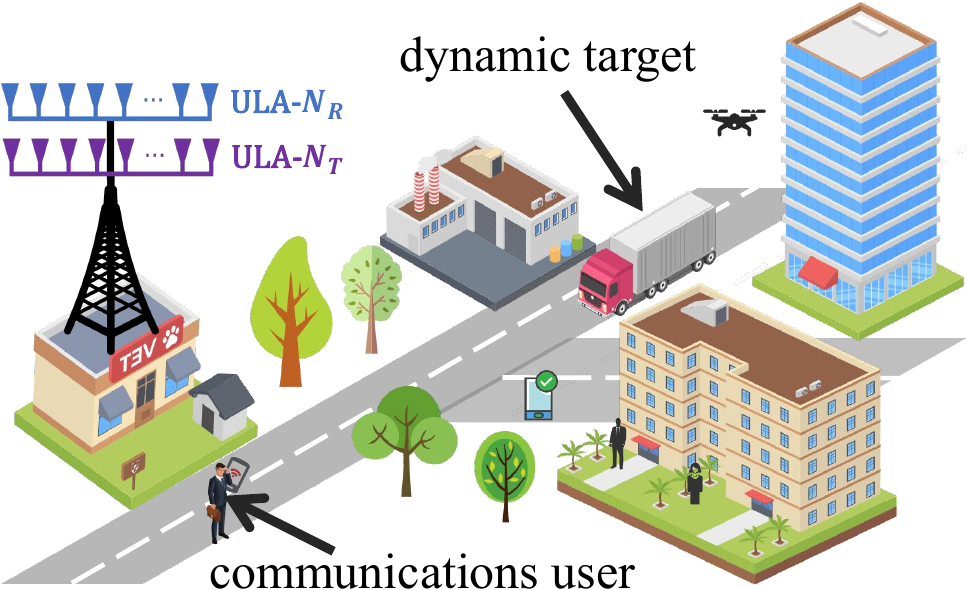}
	\caption{System model.}
	\label{fig_1}
\end{figure}

\emph{Notation}:
Lower-case and upper-case boldface letters $\mathbf{a}$ and $\mathbf{A}$ denote a vector and a matrix;
$\mathbf{a}^T$ and $\mathbf{a}^H$ denote the transpose and the conjugate transpose of vector $\mathbf{a}$, respectively;
$\mathbf{a}[n]$  denotes the $n$-th element of the vector $\mathbf{a}$;
$\mathbf{A}[i,j]$ denotes the $(i,j)$-th element of the matrix $\mathbf{A}$; $\mathbf{A}[i_1:i_2,:]$ is the submatrix composed of all columns elements in rows $i_1$ to $i_2$ of matrix $\mathbf{A}$;
$\mathbf{A}[:,j_1:j_2]$ is the submatrix composed of all rows elements in columns $j_1$ to $j_2$ of matrix $\mathbf{A}$;
$\left|\cdot\right|$  denotes the absolute operator;
${\rm eig}(\cdot)$ represents the matrix eigenvalue decomposition function.
$\mathbb{R}$ and $\mathbb{C}$  represent the  real field and complex field, respectively.
$\mathbb E\{ \cdot \}$ is the mathematical expectation operator.

\section{System Model and Proposed ISAC Framework}

In this section, we will  propose a practical ISAC framework that effectively integrates the sensing functions into the existing communications systems.

\subsection{BS Model}

\begin{figure*}[!t]
	\centering
	\includegraphics[width=135mm]{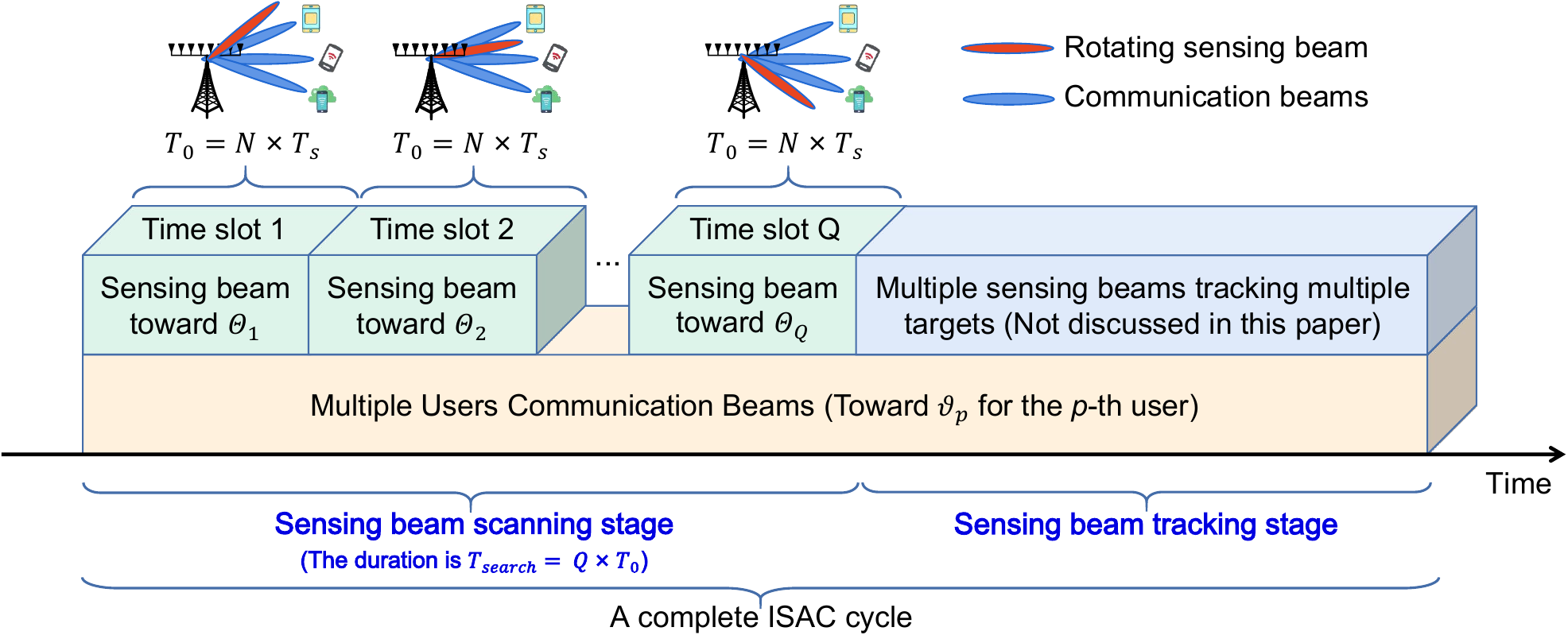}
	\caption{The proposed ISAC framework.}
	\label{fig_1}
\end{figure*}

A massive MIMO based ISAC system
operating in  mmWave or Terahertz frequency bands
  with OFDM modulation is depicted in Fig.~2, which employs a dual function BS for wireless communication and radar sensing at the same time.
The BS is configured with two parallel and closely placed  uniform linear arrays (ULAs) of $N_T$ and $N_R$ antenna elements  as  transmitting array and  receiving array, respectively, in which  the antenna spacing  is $d\le \frac{\lambda}{2}$ and $\lambda$ is the wavelength. 
Assuming that the ULAs  are parallel to the y-axis, and  
the antenna indices in the transmitting- and receiving- ULAs are $0,...,N_T-1$ and $0,...,N_R-1$.
Suppose that the communications system uses \textcolor{black}{narrowband}   OFDM signals with $M$ subcarriers in total,
and the  lowest  frequency and the subcarrier  interval are $f_0$ and $\Delta f$, respectively.
 Then the transmission bandwidth is $W=(M-1)\Delta f$, and 
 the frequency of the $m$-th subcarrier  is $f_m=f_0+m\Delta f$, where $m=0,1,...,M-1$.

Assume that the service area of   BS is
$\{(r,\theta)|r_{min}\leq r \leq r_{max},\theta_{min}\leq \theta \leq \theta_{max}\}$,  
and there are $P$ single-antenna communications users, $K$ dynamic targets, as well as widely distributed static environment within this  area.
We assume that the position and radial velocity of the $k$-th dynamic target are $(r_k,\theta_k)$ and $v_k$.
Besides,
we  assume that  the position of the $p$-th user is $(R_p,\vartheta_p)$,
and suppose that the positions of   users are known and are stationary to  BS, as the cooperative users' positions can  be easily obtained through  user reporting, 
or other techniques\cite{9782674,10271123,9013639}.

\subsection{Proposed ISAC Framework}


The task of an ISAC system is to sense all $K$ dynamic targets while serving the communications of all $P$ users, in which the  sensing of dynamic targets  includes detection, estimation, and tracking.
As described in Fig.~3,
the proposed ISAC framework consists of two stages: \emph{sensing beam scanning (SBS) stage} and  \emph{sensing beam tracking (SBT) stage}.
For the aspect of communications,   BS continuously generates $P$ communications beams  towards $P$    users
to  maintain  communications service during both SBS stage and SBT stage. 

For the aspect of sensing,  BS 
generates one  sensing beam that can rotate and scan 
the  service area during  SBS stage, that is,  the sensing beam can explore all angle spaces at certain  scanning intervals within a certain period of time,
and  BS can detect the  targets
and estimate their parameters during  SBS stage.
Next,  BS should generate one or $K$ sensing beams  to track all $K$  dynamic targets 
during  SBT stage.
In this work, we  mainly focuses on   SBS stage,
while  SBT stage is  a well separate issue and will be discussed in subsequent works.

Suppose that  BS adopts $N$ consecutive OFDM symbols to realize dynamic target sensing in one single direction, and  OFDM symbol interval time is  $T_s = \frac{1}{\Delta f}$.
As shown in Fig.~3,
we divide the SBS stage  into $Q$ transient time slots,  
and  each time slot lasts for a time duration of $T_0 = N\cdot T_s$.
In the $q$-th time slot,   BS needs to generate $P$ communications beams pointing to $P$ users and  generate one sensing beam pointing to the \emph{sensing scanning angle} $\Theta_q$ from  transmitting array.
\textcolor{black}{
Here we assume that the transmission power of  BS is $P_t$,
the energy of sensing beam is  $\rho_{q}^s P_t$ with $\rho_{q}^s\in [0,1]$,
the energy of  communications beam for the 
$p$-th user is $\rho_{q,p}^c P_t$ with $\rho_{q,p}^c \in [0,1]$, where $p=1,2,...,P$,
and there is $\rho_{q}^s + \sum_{p=1}^{P} \rho_{q,p}^c =1$. We collectively refer to $\rho_{q}^s, \rho_{q,1}^c, \rho_{q,2,}^c,...,\rho_{q,P}^c$ as the \emph{power allocation factor}.}
Then the transmission signals  on the $m$-th subcarrier of the $n$-th OFDM symbol in the $q$-th time slot should be represented as
\begin{equation}
\begin{split}\color{black}
\begin{aligned}
\label{deqn_ex1a}
& \mathbf{x}_{q,n,m} = \sum_{p=1}^{P}\mathbf{w}_{c,p,q}s^{c,p,q}_{n,m}+\mathbf{w}_{s,q}s^{s,q}_{n,m}\\&
= \!\! \sum_{p=1}^{P}\!\sqrt{\!\frac{\rho_{q,p}^c P_t}{N_T}}\mathbf{a}_{T\!X}(\vartheta_p)s^{c,p,q}_{n,m}\!+\!\!\sqrt{\!\frac{\rho_q^s P_t}{N_T}}\mathbf{a}_{T\!X}(\Theta_q\!)s^{s,q}_{n,m},
\end{aligned}
\end{split}
\end{equation}
\textcolor{black}{where $\mathbf{w}_{c,p,q}=\sqrt{\!\frac{\rho_{q,p}^c  \!P_t}{N_T}}\mathbf{a}_{T\!X}(\vartheta_p)$ and $\mathbf{w}_{s,q}=\sqrt{\frac{\rho_q^s P_t}{N_T}}\mathbf{a}_{T\!X}(\Theta_q)$} are the communications beamforming vector for the $p$-th user  and the sensing beamforming vector for the  $\Theta_q$, respectively, and $\mathbf{a}_{TX}(\theta)$ is the transmitting  steering vector for  angle $\theta$ with 
\begin{equation}
\begin{split}
\begin{aligned}
\label{deqn_ex1a}
\mathbf{a}_{T\!X}(\theta)\! =\! [1, e^{j2\pi\! f_0\!\frac{d\sin\theta}{c}},...,e^{j2\pi\! f_0\!\frac{(N_T\!-\!1)d\sin\theta}{c}}]^T \!\in\! \mathbb{C}^{N_T\times 1}.
\end{aligned}
\end{split}
\end{equation}
 Moreover,  $s^{c,p,q}_{n,m}$ and $s^{s,q}_{n,m}$ are the communication signals for the $p$-th user and the 
sensing detection signals transmitted by  BS  through the $m$-th subcarrier of the $n$-th OFDM symbol in the $q$-th time slot.
Without loss of generality, $s^{c,p,q}_{n,m}$ and $s^{s,q}_{n,m}$  are both assumed as zero-mean,   temporally-white and  wide-sense stationary stochastic process, and  $s^{c,p,q}_{n,m}$ is uncorrelated with $s^{s,q}_{n,m}$.  Here, both  $s^{c,p,q}_{n,m}$ and $s^{s,q}_{n,m}$ are normalized to have unit power, and the actual transmission power will be calculated in the corresponding beamforming vectors.

In addition, to  suppress the interference of echoes from other directions,
  BS also demands to set one sensing receiving beam pointing to $\Theta_q$ at  receiving array.
Here
we  set the receiving beamforming vector in the $q$-th time slot as $\mathbf{w}_{RX,q}=\frac{1}{\sqrt{N_R}}\mathbf{a}_{RX}(\Theta_q)$, where   $\mathbf{a}_{RX}(\theta)$ is the receiving  steering vector for  angle $\theta$ with the form
\begin{equation}
\begin{split}
\begin{aligned}
\label{deqn_ex1a}
\mathbf{a}_{R\!X}\!(\theta) = [1, e^{j2\pi \!f_0\!\frac{d\sin\theta}{c}},...,e^{j2\pi \!f_0\!\frac{(N_R\!-\!1)d\sin\theta}{c}}]^T \!\in\! \mathbb{C}^{N_R\!\times \!1}.
\end{aligned}
\end{split}
\end{equation}

To evenly scan the  service area, we  set 
$\Theta_q = \arcsin [\sin\theta_{min}+(q-1)\frac{\sin\theta_{max}-\sin\theta_{min}}{Q-1}]$, where $q=1,2,...,Q$.
\textcolor{black}{Therefore,  the total duration of beam scanning in  SBS stage can be expressed as $T_{SBS} = Q\times N\times T_s$.\footnote{\textcolor{black}{Generally, the overall sensing accuracy will improve with the increase of beam  scanning time, and one can adjust the beam scanning time according to the required sensing accuracy.}}} 
Then  BS will transmit signals through  transmitting array and receive echoes through  receiving array during each time slot.
After completing the beam scanning, we need to design
efficient and accurate dynamic target sensing algorithms to detect
the dynamic targets and to estimate their angles, distances, and velocities 
 in clutter environment.

\subsection{Interference Management}

During SBS stage, due to the  rotation of  sensing beam, the intervals between the direction of  sensing beam and  directions of  communications beams are constantly changing.
Hence, sensing beam would bring dynamic interference to communications beams.
Specifically, when the sensing scanning angle $\Theta_q$ is far from  user's angle, the interference of  sensing beam on  communications beams is relatively small.
However,  when $\Theta_q$ is close to  user's angle, the interference of  sensing beam on  communications beams is significant large.
Especially when the mainlobe of  sensing beam overlaps with the mainlobe of  communications beam,  sensing beam will cause serious interference to this communications beam.
Hence we should properly control the power allocated to sensing beam \textcolor{black}{and communications beams}  to suppress   dynamic interference.

\begin{figure}[!t]
\centering
\includegraphics[width=89mm]{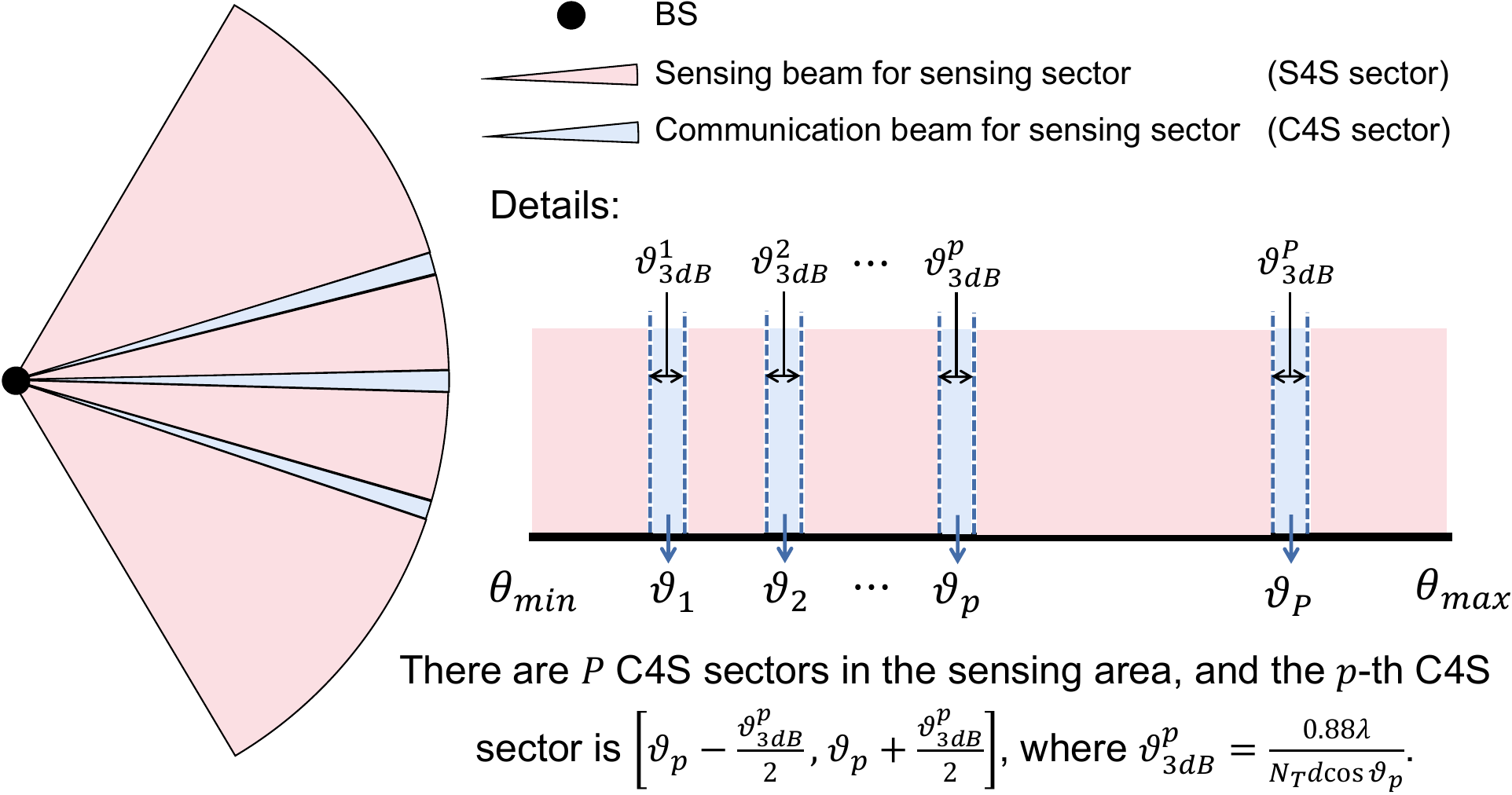}
\caption{Schematic diagram of S4S sector and C4S sector.}
\label{fig_1}
\end{figure}

Intuitively,  when  $\Theta_q$ is close to  $\vartheta_p$, in order to avoid serious interference caused by  sensing beam on  user, BS should not generate a dedicated sensing beam at this time.
\textcolor{black}{Interestingly, 
as the communications beam for the user can also illuminate the
potential targets that may exist in this direction,  the communications beam  would also  cause  echoes  if  there are some targets in this direction,
and 
the BS can still utilize the echoes of communications beam itself to sense the targets.}
Hence we divide the  service area of  BS  into  \emph{sensing beam for sensing (S4S) sectors}
and  \emph{communications beam for sensing (C4S) sectors} as shown in Fig.~4.
In general, we construct a protective C4S sector for each  user, and the C4S sector corresponding to the $p$-th user is
 $[\vartheta_p-\frac{\vartheta^p_{3dB}}{2},\vartheta_p+\frac{\vartheta^p_{3dB}}{2}]$,
where $\vartheta^p_{3dB} = \frac{0.88\lambda}{N_Td \cos\vartheta_p}$ is the half power beam width\footnote{This paper takes $\vartheta^p_{3dB}$ as  a typical value for the width of the C4S sector, which can be adjusted according to actual performance in practical  systems.}.
Then we should take the following strategies to suppress the interference of sensing beam on communications beams:
\begin{enumerate}
\item When  $\Theta_q$ is located within the $p$-th C4S sector,  we design that the  BS does not generate the dedicated sensing beam, but utilizes the communications beam toward   $\vartheta_p$  to illuminate the potential targets  and  realize  sensing.
\textcolor{black}{
In this case, 
the BS only needs to optimize the power allocation of $P$ communication beams, that is,
the BS only needs to optimize the power allocation factor $\{\rho_{q,1}^c,\rho_{q,2}^c,...,\rho_{q,P}^c\}$, while $\rho_q^s$ is fixed as  $\rho_q^s = 0$.}

\item When $\Theta_q$ is located within  S4S sectors,
the BS needs to optimize $P$ communications beams and one sensing beam simultaneously 
to maximize the  sensing performance while 
ensuring   communications performance.
\textcolor{black}{That is, the BS  needs to optimize the power allocation factor $\{\rho_{q}^s, \rho_{q,1}^c,\rho_{q,2}^c,...,\rho_{q,P}^c\}$.}

\end{enumerate}
The difficulty of the above  design is  that the 
 interference of rotating sensing beam on  communications beams  is dynamically changing during SBS stage.
Therefore,  BS must carefully and quickly update  power allocation factor for each time slot to suppress  interference,
which will be addressed in Section~\RNum{3}.

\subsection{Clutter Environment}

Note that the receiving array  will receive both the effective echoes caused by interested dynamic targets (dynamic target echoes) and the undesired  echoes caused by uninterested background environment.
In radar systems, these undesired  echoes   are  usually referred to as ``clutter'', including ground clutter, sea clutter,  weather clutter,  birds clutter, etc\cite{784056,766939}. 
The key difference between dynamic target echoes and clutter lies in their different Doppler frequencies.
That is, the Doppler frequency of dynamic target echoes is usually much higher than that of clutter, while the Doppler frequency of clutter is usually zero or a small non-zero value.
Specifically, the 
 ground clutter is usually caused by  static environment, such as land, mountains, roads, and buildings, etc,
 and its signal power intensity is usually  much higher than that of the dynamic target echo, but its Doppler frequency  is almost zero\cite{1457401,skolnik1970radar}.
Hence a practical modeling of echo signals should include both dynamic targets echoes and static environment echoes, and  BS needs to 
 accurately detect the  dynamic targets and estimate their parameters   under   clutter environment.
However, 
most existing ISAC works \cite{9898900,10050406,8918315,8114253,10048770} ignore  the  interference caused by static environmental   clutter on dynamic target sensing, 
 which does not match  the real scenarios.
We will construct a practical ISAC scene with clutter environment and provide a complete dynamic targets sensing scheme in Section~\RNum{4}.

\section{Joint Communications and Sensing  Power Allocation Optimization}

In this section, we will jointly optimize ISAC
power allocation for each time slot.

\subsection{Communications Performance Metric}
Let us first describe the communications process.
The  communications channel of the $p$-th   user on the $m$-th subcarrier of the $n$-th OFDM symbol can be represented as
\begin{equation}
\begin{split}
\begin{aligned}
\label{deqn_ex1a}
\mathbf{h}_{c,p,n,m} = \gamma_p e^{-j2\pi f_m\frac{R_p}{c}}\mathbf{a}_{TX}(\vartheta_p) \in \mathbb{C}^{N_T\times 1},
\end{aligned}
\end{split}
\end{equation}
where $\gamma_p = \sqrt{\frac{\lambda^2}
{(4\pi R_p)^2}}$ is the channel fading for the $p$-th user.
Then the received communications signal of the $p^*$-th user on the $m$-th subcarrier of the $n$-th  symbol in the $q$-th time slot is $y^{c,p^*,q}_{n,m}$, which is
 expressed as  (5) at the top of  this  page,
where  $\gamma_{p^*}' \triangleq \gamma_{p^*} e^{j2\pi f_m\frac{R_{p^*}}{c}}$, 
 $n^{c,p^*,q}_{n,m}$  is the zero-mean additive complex  Gaussian
    white  noise at the user with variance  $\sigma_{c,p^*}^2$,
and
$F_{TX}(\theta_1,\theta_2) \triangleq \mathbf{a}_{TX}^H(\theta_{1})\mathbf{a}_{TX}(\theta_2)$.

\newcounter{TempEqCnt} 
\setcounter{TempEqCnt}{\value{equation}} 
\setcounter{equation}{4} 
\begin{figure*}[ht] 
\begin{equation}
\begin{split}\color{black}
\begin{aligned}
\label{deqn_ex1a}
y^{c,p^*,q}_{n,m} &= \mathbf{h}_{c,p^*,n,m}^H \mathbf{x}_{q,n,m} + n^{c,p^*,q}_{n,m} =\gamma_{p^*}' \!\!
\left[\sum_{p=1}^P \sqrt{\frac{\rho_{q,p}^c P_t}{N_T}} F_{TX}(\vartheta_{p^*},\vartheta_p)s^{c,p,q}_{n,m} \!+\! \sqrt{\frac{\rho_q^s P_t}{N_T}}F_{TX}(\vartheta_{p^*},\Theta_q)s^{s,q}_{n,m} 
\right]\! +\! n^{c,p^*,q}_{n,m}\\
&=\underbrace{\gamma_{p^*}'\!\sqrt{\!{\rho_{q,p*}^c P_tN_T}}s^{c,p^*,q}_{n,m}}_{{\rm ER}} \!+\! \underbrace{\gamma_{p^*}'\!\sum_{p=1,p\neq p^*}^P \sqrt{\!\frac{\rho_{q,p}^c P_t}{N_T}} \!F_{TX}(\vartheta_{p^*},\vartheta_p)s^{c,p,q}_{n,m}}_{{\rm MUI}}+\underbrace{\gamma_{p^*}'
\sqrt{\frac{\rho_q^sP_t}{N_T}}F_{TX}(\vartheta_{p^*},\Theta_q)s^{s,q}_{n,m}}_{{\rm SI}}  + \underbrace{n^{c,p^*,q}_{n,m}}_{{\rm Noise}}.
\end{aligned}
\end{split}
\end{equation}
\hrulefill 
\vspace*{8pt} 
\end{figure*}

Eq. (5) indicates that the  received communications signal is composed of four parts: effective reception (ER), multiple users interference (MUI), sensing interference (SI), and noise (Noise).
Note that the sensing beam has a negative interfering effect on the communications beams,
and thus the signal-to-interference-plus-noise ratio (SINR) should be used to describe the  communications performance. Specifically,
 the SINR for the $p^*$-th user in the $q$-th time slot is
\begin{equation}
\begin{split}\color{black}
\begin{aligned}
\label{deqn_ex1a}
 &{\rm SINR}_{c,p^*,q} = \frac{\mathbb E\{|{\rm ER}|^2\}}{\mathbb E\{|{\rm MUI}|^2\}+\mathbb E\{|{\rm SI}|^2\}+\mathbb E\{|{\rm Noise}|^2\}}\\
&=\frac{\rho_{q,p*}^cN_T^2}
{\sum_{p\!=\!1,p\neq p^*}^P \! \rho_{q,p}^c G_{T\!\!X}\!(\!\vartheta_{p^*}\!,\!\vartheta_p\!)\! \!+\!\! \rho_q^s G_{T\!\!X}\!(\!\vartheta_{p^*}\!,\!\Theta_q\!)\!\! +\!\! \frac{N_T\sigma_{c,p^*}^2}{P_t\gamma_{p^*}^2} },
\end{aligned}
\end{split}
\end{equation}
where $\mathbb E\{ \cdot \}$ is the mathematical expectation operator
and $G_{TX}(\theta_1,\theta_2)$ is defined as
\begin{equation}
\begin{split}
\begin{aligned}
\label{deqn_ex1a}
G_{T\!X}\!(\theta_1,\!\theta_2) \!\!=\!\! \left|\!F_{T\!X}(\theta_1,\!\theta_2)\right|^2
\!\!=\!\!\left| \!
\frac{\sin \![\!\frac{\pi df_0}{c}\!(\sin\!\theta_1 \!\!-\!\sin\!\theta_2)\!N_T]}
{\sin[\frac{\pi df_0}{c}(\sin\theta_1\!-\!\sin\theta_2)]}\!
\right|^2\!\!\!.
\end{aligned}
\end{split}
\end{equation}

\subsection{Sensing Performance Metric}

\textcolor{black}{The existing ISAC waveform optimization studies usually adopt the  signal-to-noise ratio (SNR) of echo signals, and
Cramer-Rao lower bound (CRLB), etc, as the 
sensing performance metric.
These indicators are  directly related to the target's parameters information.
However, in  SBS stage, when the BS scans the angle direction $\Theta_q$, it is not known beforehand whether there is a target in this direction and the specific parameters of the target.  Hence these indicators mentioned above are not applicable to  SBS stage.}

\textcolor{black}{Note that when the BS needs to scan the sensing angle direction $\Theta_q$, the potential sensing performance will improve with the increase of \emph{equivalent transmission power} in the $\Theta_q$ direction, where the equivalent transmission power couples the actual transmission power, the array gain of the transmitting array, and the array gain of the receiving array.
We refer to the equivalent transmission power in the $\Theta_q$ direction as the \emph{equivalent sensing power (ESP)} in the $q$-th time slot,
which can be denoted as}
\begin{equation}
	\begin{split}\color{black}
		\begin{aligned}
			\label{deqn_ex1a}
\breve{P}_{s,q} = 
\mathbb E\{|\mathbf{w}_{RX,q}^H \mathbf{a}_{RX}(\Theta_q)\mathbf{a}_{TX}^H(\Theta_q)\mathbf{x}_{q,n,m}|^2\}.
		\end{aligned}
	\end{split}
\end{equation}
\textcolor{black}{Eq. (8) indicates that ESP directly couples the design of  power allocation factors, and it   effectively couples the design of antenna array pattern.
Furthermore, it should be pointed out that
ESP used in this work and the sensing performance indicators used in  \cite{8386661} and \cite{9933894} are essentially consistent.}

\textcolor{black}{By submitting (1) into (8), the ESP can be rewritten as}
\begin{equation}
	\begin{split}\color{black}
		\begin{aligned}
			\label{deqn_ex1a}
\breve{P}_{s,q} = 
\underbrace{\sum_{p=1}^{P}\frac{\rho_{q,p}^cP_tN_RG_{TX}(\Theta_q,\vartheta_p)}{N_T}}_{\rm CBE-ESP} +
\underbrace{\rho_q^sP_tN_T N_R}_{\rm SBE-ESP},
		\end{aligned}
	\end{split}
\end{equation}
\textcolor{black}{where ${\rm CBE}$-${\rm ESP}$ is the ESP contributed by $P$ communications beams, while  ${\rm SBE}$-${\rm ESP}$ is the ESP contributed by the single sensing beam.
Both the ${\rm CBE}$-${\rm ESP}$ and  ${\rm SBE}$-${\rm ESP}$ are beneficial for sensing and can be utilized by  BS for   sensing.}

\begin{figure*}[!t]
	\centering
	\subfloat[]{\includegraphics[width=45mm]{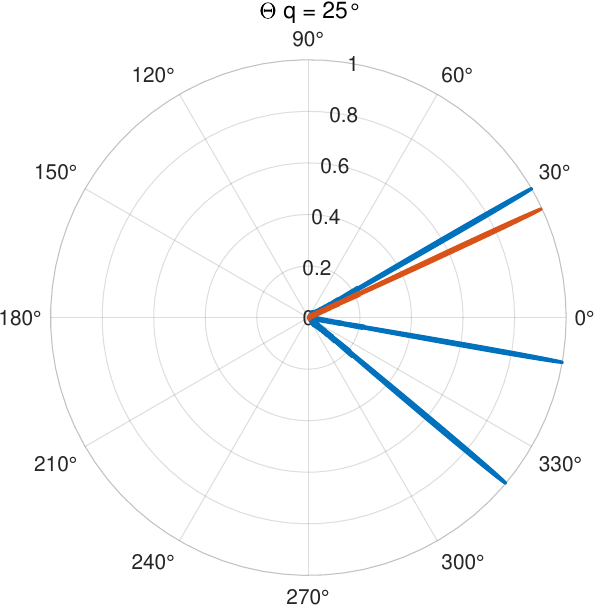}%
		\label{fig_first_case}}
	\hfil
	\subfloat[]{\includegraphics[width=45mm]{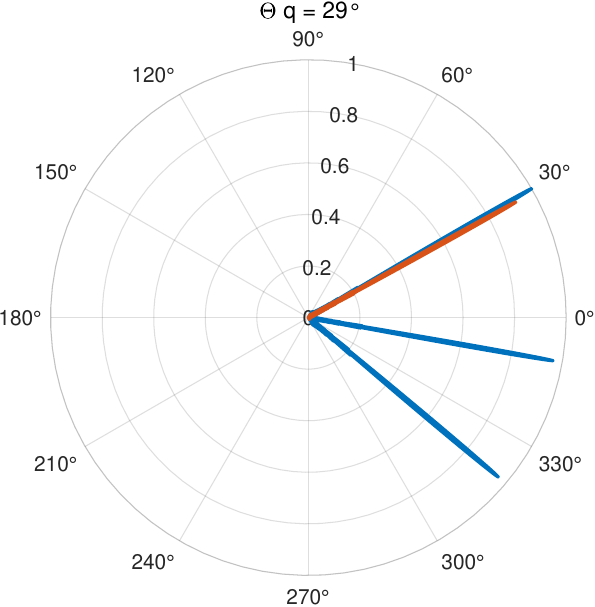}%
		\label{fig_first_case}}
	\hfil
	\subfloat[]{\includegraphics[width=45mm]{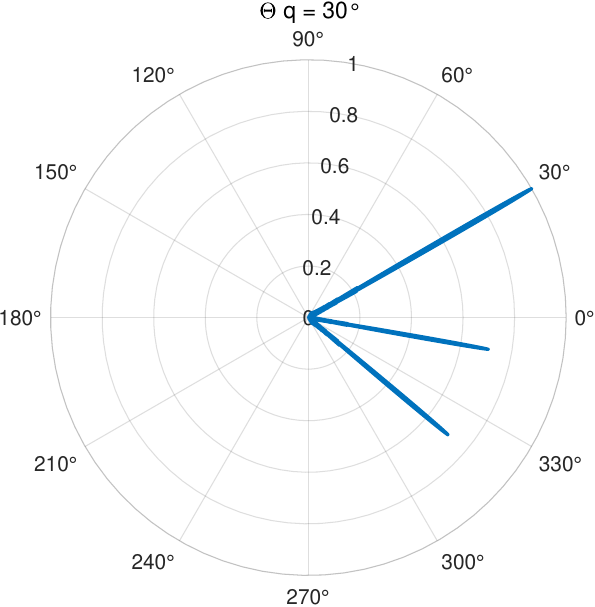}%
		\label{fig_first_case}}
	\hfil
	\subfloat[]{\includegraphics[width=45mm]{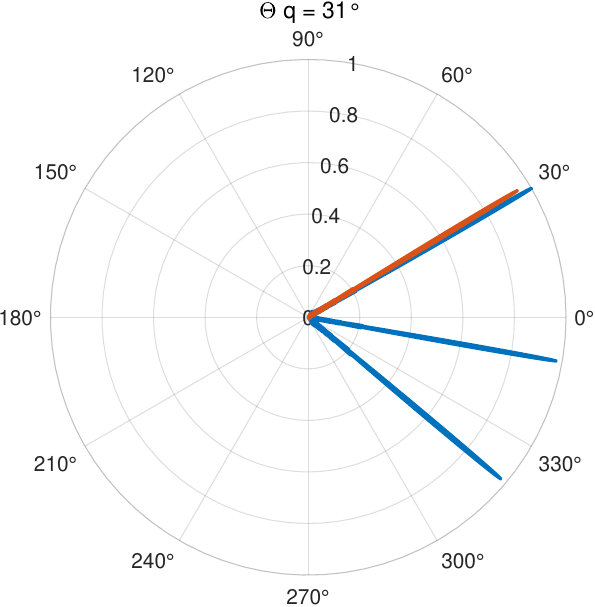}%
		\label{fig_second_case}}
	\caption{(a) An example of beamforming optimization results, where  $\Theta_q$ is $25^\circ$.
		(b) $\Theta_q = 29^\circ$.
		(c) $\Theta_q = 30^\circ$.
		(d) $\Theta_q = 31^\circ$.
		Three  users are located in the directions of $-40^\circ$, $-10^\circ$, and $30^\circ$, respectively. The blue and red lines  represent the communications beams and sensing beam, respectively.}
	\label{fig_sim}
\end{figure*}

\subsection{Joint Communications and Sensing  Power Allocation}

\textcolor{black}{
As described in Section~\RNum{2}.~C, 
when the sensing scanning angle $\Theta_q$ is located within the $p^\#$-th C4S sector,  BS no longer generates the dedicated sensing beam,
which means that we fix $\rho_q^s$ as $\rho_q^s=0$ at this time. 
Then in order to maximize  system sensing performance while 
ensure  user communications performance,
the overall optimization problem should be  formulated as}
\begin{align}\color{black}
\rm (P1): \ &\color{black} \max\limits_{\rho_{q,1}^c,...,\rho_{q,p^\#}^c,...,\rho_{q,P}^c} \quad  \breve{P}_{s,q},\label{13}\\
&\color{black} s.t. \quad   {\rm SINR}_{c,p^*,q}\geq \epsilon_{p^*}, p^* = 1,2,...,P, \label{14}\\
&\color{black} \quad\quad\ \rho_{q}^s + \sum_{p=1}^{P} \rho_{q,p}^c =1,\label{15}\\
&\color{black} \quad\quad\ \rho_{q}^s = 0,\label{16}\\
&\color{black} \quad\quad\ \rho_{q,p}^c \geq 0, p=1,2,...,P,\label{17}
\end{align}
\textcolor{black}{where $\epsilon_{p^*}$ is the minimum SINR required for the $p^*$-th user, and 
$p^\#$ is a fixed value.
To equivalently represent (P1),
let us define 
$\bm{\rho}_{q,p^\#}^{C4S} = [\rho_{q,1}^c,...,\rho_{q,p^\#}^c,...,\rho_{q,P}^c]^T\in \mathbb{R}^{P\times 1}$,
$\mathbf{n}_{q,p^\#}^{C4S}\in \mathbb{R}^{P\times 1}$ with 
$\mathbf{n}_{q,p^\#}^{C4S}[p] = 
\epsilon_{p^*}\frac{N_T\sigma_{c,p^*}^2}{P_t\gamma_{p^*}^2}$,
$\mathbf{G}_{q,p^\#}^{C4S}\in \mathbb{R}^{P\times P}$ with
$\mathbf{G}_{q,p^\#}^{C4S}[p^*,p]=G_{TX}(\vartheta_{p^*},\vartheta_p)\epsilon_{p^*}$ for $p^*\neq p$, 
$\mathbf{G}_{q,p^\#}^{C4S}[p^*,p]=-N_T^2$ for $p^* =  p$, and 
$\mathbf{m}_{q,p^\#}^{C4S}\in \mathbb{R}^{P\times 1}$ with
$\mathbf{m}_{q,p^\#}^{C4S}[p]=
\frac{P_tN_RG_{TX}(\Theta_q,\vartheta_p)}{N_T}$.
Then by submitting (6) and (9) into (11) and (10),   (P1)  can be equivalently transformed as}
\begin{align}\color{black}
	\rm (P1.1): \ &\color{black} \mathop{\rm maximize}\limits_{\bm{\rho}_{q,p^\#}^{C4S}} \quad  (\mathbf{m}_{q,p^\#}^{C4S})^T \bm{\rho}_{q,p^\#}^{C4S},\label{13}\\
	&\color{black} s.t. \quad 
\mathbf{G}_{q,p^\#}^{C4S} \bm{\rho}_{q,p^\#}^{C4S}+\mathbf{n}_{q,p^\#}^{C4S}\leq 0, \label{14}\\
	&\color{black} \quad\quad\ \mathbf{1}^T\bm{\rho}_{q,p^\#}^{C4S}-1 = 0,\label{15}\\
	&\color{black} \quad\quad\ \rho_{q}^s = 0,\label{16}\\
	&\color{black} \quad\quad\ \bm{\rho}_{q,p^\#}^{C4S} \geq \mathbf{0}. \label{17}
\end{align}
\textcolor{black}{It is seen that (P1.1) is a linear programming (LP) problem, and thus we can directly use CVX to solve it\cite{cvx,luo2022secure}.}

\textcolor{black}{On the other side,
when   $\Theta_q$ is located within S4S sectors, 
the overall optimization problem is formulated into}
\begin{align}\color{black}
	\rm (P2): \  &\color{black} \mathop{\rm maximize}\limits_{\rho_q^s,\rho_{q,1}^c,...,\rho_{q,P}^c} \quad  \breve{P}_{s,q},\label{13}\\
	&\color{black} s.t. \quad   {\rm SINR}_{c,p^*,q}\geq \epsilon_{p^*}, p^* = 1,2,...,P, \label{14}\\
	& \color{black} \quad\quad\ \rho_{q}^s + \sum_{p=1}^{P} \rho_{q,p}^c =1\label{15},\\
	& \color{black} \quad\quad\ \rho_{q}^s \geq 0, \rho_{q,p}^c\geq 0, p=1,...,P\label{16}.
\end{align}
\textcolor{black}{Let us define 
$\bm{\rho}_{q}^{S4S} = [\rho_{q,1}^c,...,\rho_{q,P}^c,\rho_q^s]^T\in \mathbb{R}^{(P+1)\times 1}$,
$\mathbf{n}_{q}^{S4S}\in \mathbb{R}^{P\times 1}$ with 
$\mathbf{n}_{q}^{S4S}[p] = 
\epsilon_{p^*}\frac{N_T\sigma_{c,p^*}^2}{P_t\gamma_{p^*}^2}$,
$\mathbf{m}_{q}^{S4S}\in \mathbb{R}^{(P+1)\times 1}$ with
$\mathbf{m}_{q}^{S4S}[p]=
\frac{P_tN_RG_{TX}(\Theta_q,\vartheta_p)}{N_T}$ for $p=1,...,P$,
$\mathbf{m}_{q}^{S4S}[P+1]=P_tN_TN_R$, and $\mathbf{G}_{q}^{S4S}\in \mathbb{R}^{P\times (P+1)}$ with}
\begin{equation}
	\begin{split}\color{black}
		\begin{aligned}
			\label{deqn_ex1a}
\mathbf{G}_{q}^{S4S}[p^*,p] =
			\left\{ \!\!\!\!\!\!\!\!
			\begin{array}{rcl}
&G_{TX}(\vartheta_{p^*},\vartheta_p)\epsilon_{p^*},& 
 {\rm if} \ p^* \neq p, 1\leq p\leq P\\
 &-N_T^2,& 
 {\rm if} \ p^* = p, 1\leq p\leq P\\
&G_{TX}(\vartheta_{p^*},\Theta_q)\epsilon_{p^*},& {\rm if} \ p=P+1.
			\end{array} \right.
		\end{aligned}
	\end{split}
\end{equation}
\textcolor{black}{Then by submitting (6) and (9) into (21) and (20),   (P2)  can be equivalently transformed into}
\begin{align}\color{black}
	\rm (P2.1): \ &\color{black} \max\limits_{\bm{\rho}_{q}^{S4S}} \   (\mathbf{m}_{q}^{S4S})^T \bm{\rho}_{q}^{S4S},\label{13}\\
	&\color{black} s.t. \quad 
	\mathbf{G}_{q}^{S4S} \bm{\rho}_{q}^{S4S}+\mathbf{n}_{q}^{S4S}\leq 0, \label{14}\\
	&\color{black} \quad\quad\ \mathbf{1}^T\bm{\rho}_{q}^{S4S}-1 = 0,\label{15}\\
	&\color{black} \quad\quad\ \bm{\rho}_{q}^{S4S} \geq \mathbf{0}. \label{16}
\end{align}
\textcolor{black}{It can be found that (P2.1) is also an LP problem, and thus we can  use CVX to solve it too.}

Fig.~5 shows the examples of  transmitting beamforming  with multiple users, in which $N_T=128$  antennas are configured and three  users are located in   $-40^\circ$, $-10^\circ$ and $30^\circ$, respectively.
The blue and red lines  represent the communications beams and sensing beam, respectively.
 Fig.~5(a) to Fig.~5(d)  show several silhouettes  with $\Theta_q$ gradually scanning from $25^\circ$  to $31^\circ$,
and this process undergoes a C4S sector formed by  the user at $30^\circ$
as $[29.5452^\circ,30.4548^\circ]$.
In Fig.~5(a),  due to the difference between $\Theta_q = 25^\circ$  and  users angles, 
an obvious red sensing beam in  $25^\circ$ direction can be observed.
In Fig.~5(c), the $\Theta_q$  is set as  $30^\circ$, which is within one  C4S sector. Then the    optimization result at this time is $\rho_{q}^s = 0$, which means that  BS does not allocate energy to the sensing beam in order to ensure the communications performance of the  user in $30^\circ$ direction.  
Fortunately, the communications beam focused toward the $30^\circ$ direction can still be utilized by the BS for sensing at this time.

\section{Dynamic Target Detection and Estimation in Clutter Environment}

In this section, we will construct a practical ISAC scene with clutter environment and provide a complete and practical scheme 
to sense    dynamic targets from   clutter environment.

\begin{figure}[!t]
\centering
\includegraphics[width=50mm]{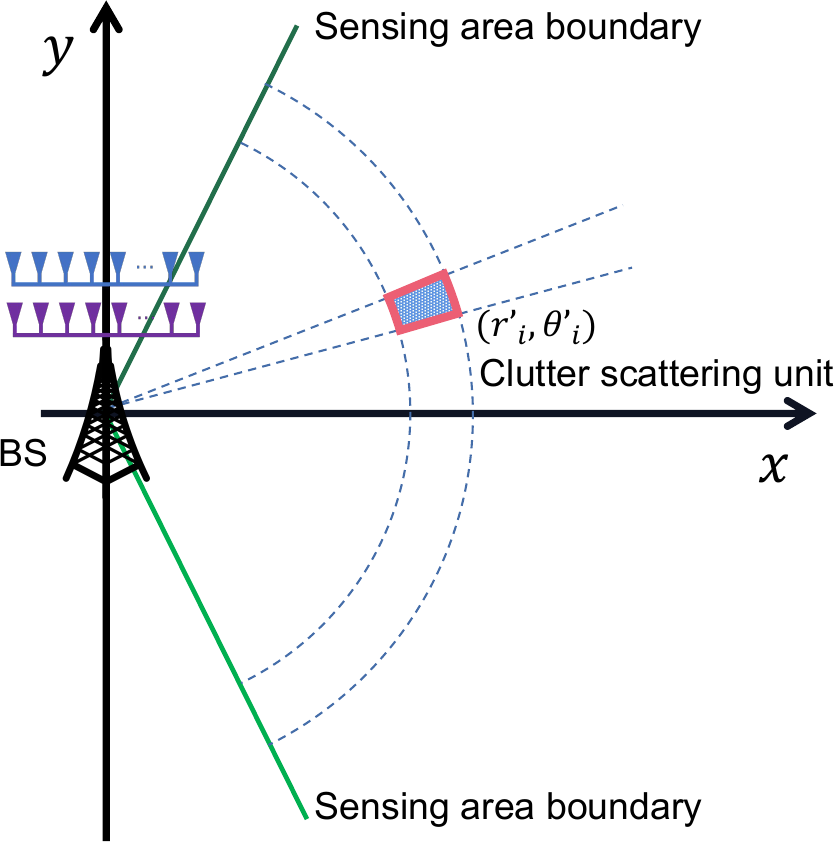}
\caption{Schematic diagram of static environmental clutter scattering unit.}
\label{fig_1}
\end{figure}

\subsection{Echo Signals Model}

As described in Section~\RNum{2}.~D, the  original echo signals should include both dynamic targets echoes (DTE) and static environment echoes (SEE). 
Based on typical urban service scenarios, we model the clutter environment as  ground clutter.
Specifically, we divide the  clutter area within  sensing range into
$I$ static clutter scattering units based on distance and angle dimensions 
as the  blue grid unit in Fig.~6, and 
the size of the clutter scattering unit is determined by the angle and distance resolution of the system.
Assuming  the center position of the $i$-th  unit is $(r_i',\theta_i')$, Then the echo channel of the $i$-th  unit on the $m$-th subcarrier of the $n$-th  symbol can be modeled as
\begin{equation}
\begin{split}
\begin{aligned}
\label{deqn_ex1a}
\mathbf{H}_{i,n,m}' \!=\! \beta_i e^{-j2\pi f_m\!\frac{2r_i'}{c}}\mathbf{a}_{R\!X}(\theta_i')\mathbf{a}_{T\!X}^H(\theta_i'),\kern 10pt i=1,...,I,
\end{aligned}
\end{split}
\end{equation}
 where  $\beta_i =\sqrt{\frac{\lambda^2}{(4\pi)^3 r_i'^4}}\sigma_{c,i}'$
 is the channel fading factor
and $\sigma_{c,i}'$ is the radar cross section (RCS) of the $i$-th   clutter scattering unit.
The RCS of ground clutter can usually be assumed to follow 
the Swerling~$\rm \uppercase\expandafter{\romannumeral1}$ model, and the probability density function of  RCS $\sigma$ satisfies
\begin{equation}
\begin{split}
\begin{aligned}
\label{deqn_ex1a}
f(\sigma) = \frac{1}{\sigma_0} \exp(-\frac{\sigma}{\sigma_0}), \sigma \geq 0,
\end{aligned}
\end{split}
\end{equation}
where $\sigma_0$ is  the average value of  object's RCS.
Besides, the echo channel of the $k$-th dynamic target on the $m$-th subcarrier of the $n$-th OFDM symbol can be represented as
\begin{equation}
\begin{split}
\begin{aligned}
\label{deqn_ex1a}
\mathbf{H}_{k,n,m} \!=\! \alpha_k e^{j2\pi f_0 \!\frac{2v_k}{c}\!nT_s} e^{-j2\pi f_m\!\frac{2r_k}{c}}\mathbf{a}_{R\!X}(\theta_k)\mathbf{a}_{T\!X}^H(\theta_k), \\
\kern 10pt
k=1,...,K,
\end{aligned}
\end{split}
\end{equation}
where $n=0,1,2,...,N-1$, 
$\alpha_k =\sqrt{\frac{\lambda^2}{(4\pi)^3 r_k^4}}\sigma_{c,k}$, and 
 $\sigma_{c,k}$  is the RCS of the $k$-th dynamic target that   also follows  Swerling $\rm \uppercase\expandafter{\romannumeral1}$ model.
Based on (29) and (31),
the overall sensing echoes channel matrix  on the $m$-th subcarrier of the $n$-th OFDM symbol can be represented as
\begin{equation}
\begin{split}
\begin{aligned}
\label{deqn_ex1a}
\mathbf{H}_{n,m} = \sum_{k=1}^K \mathbf{H}_{k,n,m} + \sum_{i=1}^I \mathbf{H}_{i,n,m}'.
\end{aligned}
\end{split}
\end{equation}
Based on  (1) and (32),
the echo signal on the $m$-th subcarrier of the $n$-th OFDM symbol received by  BS in the $q$-th time slot is
\begin{equation}
\begin{split}
\begin{aligned}
\label{deqn_ex1a}
&y^{s,q}_{n,m} \!=\! \mathbf{w}_{RX,q}^H \mathbf{H}_{n,m} \mathbf{x}_{q,n,m} + n^{s,q}_{n,m} 
\\& \!=\! \sum_{p=1}^{P}\!\mathbf{w}_{R\!X,q}^H \mathbf{H}_{n,m}\mathbf{w}_{c,p,q}s^{c,p,q}_{n,m}\!+\!\mathbf{w}_{R\!X,q}^H \mathbf{H}_{n,m}\mathbf{w}_{s,q}s^{s,q}_{n,m} \!\!+\! n^{s,q}_{n,m},
\\& \kern 10pt q=1,...,Q,\kern 4pt n=0,...,N-1,\kern 4pt m=0,...,M-1,
\end{aligned}
\end{split}
\end{equation}
where $n^{s,q}_{n,m}$ is the zero-mean additive   Gaussian   noise  with variance  $\sigma_{s,q}^2$.
Then we can stack $y^{s,q}_{n,m}$ into one echoes tensor 
${\mathbf{Y}}_{cube} \in \mathbb{C}^{Q\times N\times M}$, 
whose $(q,n,m)$-th element is ${\mathbf{Y}}_{cube}[q,n,m] = {y}^{s,q}_{n,m}$.

Note that ${\mathbf{Y}}_{cube}$ includes the echoes channel, receiving and transmitting beamforming, and transmission symbols, while   targets sensing can be understood as an estimation of  echoes channel. However, random transmission symbols would affect the estimation of  echoes channel, and thus we need to  erase the transmission symbols from the received signals to obtain the equivalent echoes channel (EEC)\cite{10048770}.
For massive MIMO system, when $\Theta_q$ is located within the S4S sector,
it can be inferred from (33) that $y^{s,q}_{n,m}\approx \mathbf{w}_{RX,q}^H \mathbf{H}_{n,m}\mathbf{w}_{s,q}s^{s,q}_{n,m} + n^{s,q}_{n,m}$,
which means that  the sensing echoes are mainly provided by the sensing beam  at this time, and  thus  
the EEC corresponding to $y^{s,q}_{n,m}$ can be obtained as $\tilde{h}^{s,q}_{n,m}= y^{s,q}_{n,m}/s^{s,q}_{n,m}$. 
When $\Theta_q$ is located within the $p$-th C4S sector,
it can be inferred from (33) that $y^{s,q}_{n,m}\approx \mathbf{w}_{RX,q}^H \mathbf{H}_{n,m}\mathbf{w}_{c,p,q}s^{c,p,q}_{n,m} + n^{s,q}_{n,m}$,
which means that 
the sensing echoes are mainly provided by the $p$-th communications beam  at this time, and thus  the EEC corresponding to $y^{s,q}_{n,m}$ can be obtained as $\tilde{h}^{s,q}_{n,m}= y^{s,q}_{n,m}/s^{c,p,q}_{n,m}$.
Then we can comprehensively define the EEC as 
$\tilde{h}^{s,q}_{n,m} = y^{s,q}_{n,m}/s^{t,q}_{n,m}$,
where $s^{t,q}_{n,m}$ is designed as
\begin{equation}
\begin{split}
\begin{aligned}
\label{deqn_ex1a}
s^{t,q}_{n,m} =
\left\{ \!\!\!\!\!\!\!\!
\begin{array}{rcl}
&s^{s,q}_{n,m}, \\ & {{\rm if} \  \Theta_q \notin (\vartheta_p \!-\!\frac{\vartheta^p_{3dB}}{2},\vartheta_p\!+\!\frac{\vartheta^p_{3dB}}{2}) \ {\rm for \  all} \  p=1,...,P;}\\
&s^{c,p,q}_{n,m},  \\& {{\rm if} \  \Theta_q \in [\vartheta_p\!-\!\frac{\vartheta^p_{3dB}}{2},\vartheta_p\!+\!\frac{\vartheta^p_{3dB}}{2}] \ {\rm for \  any} \  p=1,...,P.}
\end{array} \right.
\end{aligned}
\end{split}
\end{equation}
Besides, we can stack $\tilde{h}^{s,q}_{n,m}$ into an EEC tensor  $\tilde{\mathbf{H}}_{cube} \in \mathbb{C}^{Q\times N\times M}$, 
with $\tilde{\mathbf{H}}_{cube}[q,n,m] = \tilde{h}^{s,q}_{n,m}$.

\begin{figure*}[!b]
\normalsize
\setcounter{MYtempeqncnt}{\value{equation}}
\vspace*{4pt}
\hrulefill
\setcounter{equation}{37}
\begin{equation}
\begin{split}
\begin{aligned}
\label{deqn_ex1a}
&\check{h}_{q,n,m}= \tilde{\mathbf{H}}^{dynamic}_m[q,n] \approx
\mathbf{w}_{RX,q}^H(\sum_{k=1}^K\mathbf{H}_{k,n,m})\tilde{\mathbf{x}}_{q,n,m}\!+\! \check{n}^{s,q}_{n,m}\\
&=\sum_{k=1}^{K}\left\{\!\!\!\frac{\alpha_k e^{j\phi^{k,q}_{n,m}}}{\sqrt{N_R}}F_{RX}\!(\Theta_q,\theta_k)\!\!
\left[\!\sum_{p=1}^P\!\! \sqrt{\!\frac{\rho_{q,p}^cP_t}{N_T}}\!F_{TX}(\theta_k,\vartheta_p)\frac{s^{c,p,q}_{n,m}}{s^{t,q}_{n,m}}\!+
\!\!
\sqrt{\!\frac{\rho_q^sP_t}{N_T}}\!F_{TX}(\theta_k,\Theta_q)\frac{s^{s,q}_{n,m}}{s^{t,q}_{n,m}}\!\right]\!\!\right\} \!+\!  \check{n}^{s,q}_{n,m},
\end{aligned}
\end{split}\tag{38}
\end{equation}
\begin{equation}
\begin{split}
\begin{aligned}
\label{deqn_ex1a}
\check{h}_{q,n,m}  \approx 
&\left\{ \!\!\!\!\!\!\!\!
\begin{array}{rcl}
&\check{n}^{s,q}_{n,m},  & {{\rm if} \  \Theta_q \neq \theta_k \ {\rm for \  all} \  k=1,...,K,}\\
&\sum_{k'=1}^{K_q} \alpha_{k'} e^{j\phi^{k',q}_{n,m}}\sqrt{\rho_q^sP_tN_TN_R}+\check{n}^{s,q}_{n,m},  & {{\rm if} \  \Theta_q =\theta_k\neq \vartheta_p \ {\rm for \  part \ of} \  k} \ {\rm and \ for \ all} \ p=1,...,P, \\
&\sum_{k'=1}^{K_q} \alpha_{k'} e^{j\phi^{k',q}_{n,m}}\sqrt{\rho_{q,p}^cP_tN_TN_R}+\check{n}^{s,q}_{n,m},  & {{\rm if} \   \Theta_q =\theta_k = \vartheta_p \ {\rm for \  part \ of } \ k \ {\rm and  \ any} \  p=1,...,P.}
\end{array} \right.
\end{aligned}
\end{split}\tag{39}
\end{equation}
\setcounter{equation}{\value{MYtempeqncnt}}
\end{figure*}

\begin{figure}[!t]
	\centering
	\includegraphics[width=85mm]{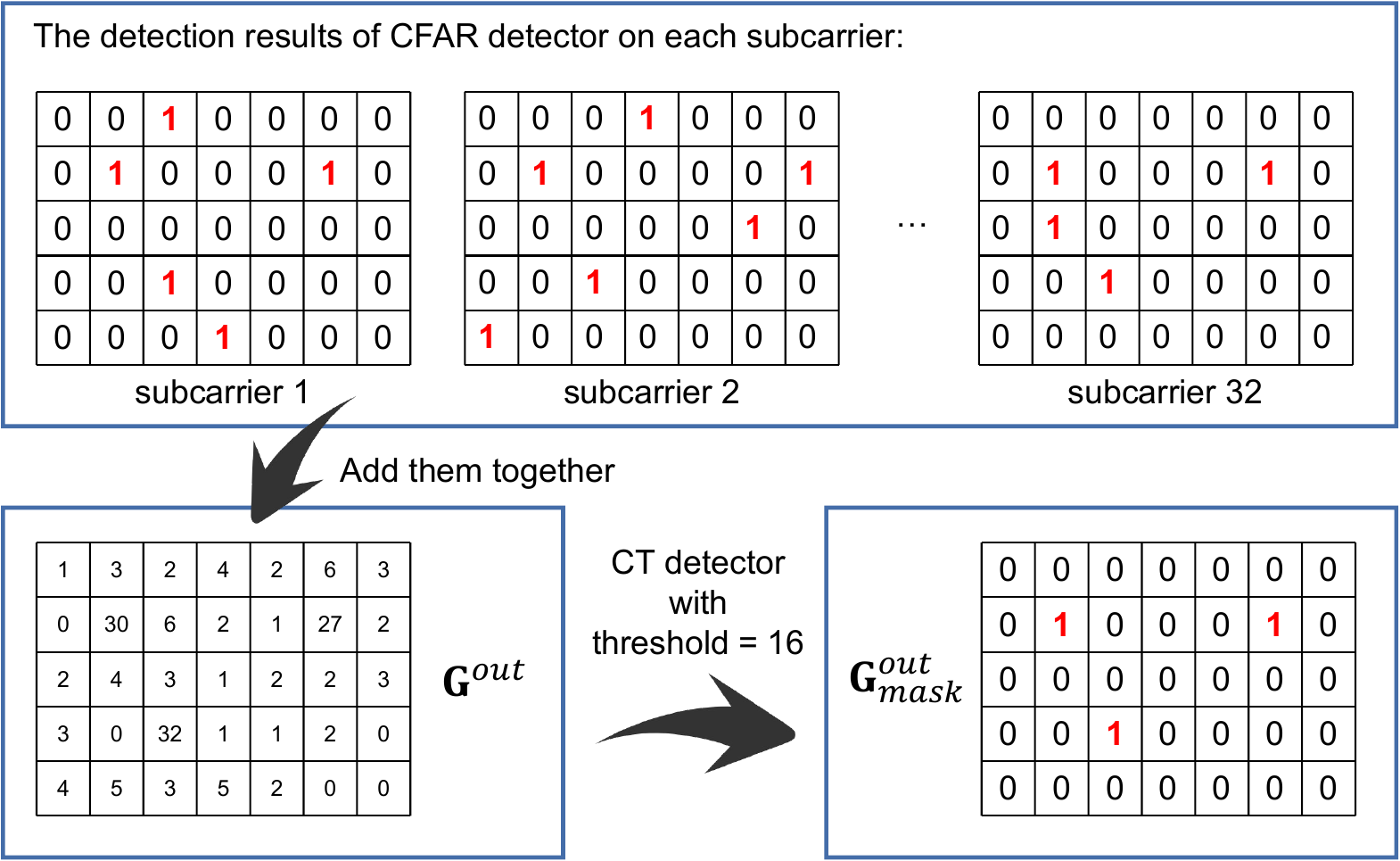}
	\caption{A simple schematic diagram of the proposed MSJD algorithm, where the number of subcarriers is $M=32$. The horizontal and vertical axes of the matrix represent angle units and velocity units, respectively.}
	\label{fig_1}
\end{figure}

\begin{figure*}[!t]
	\centering
	\subfloat[]{\includegraphics[width=70mm]{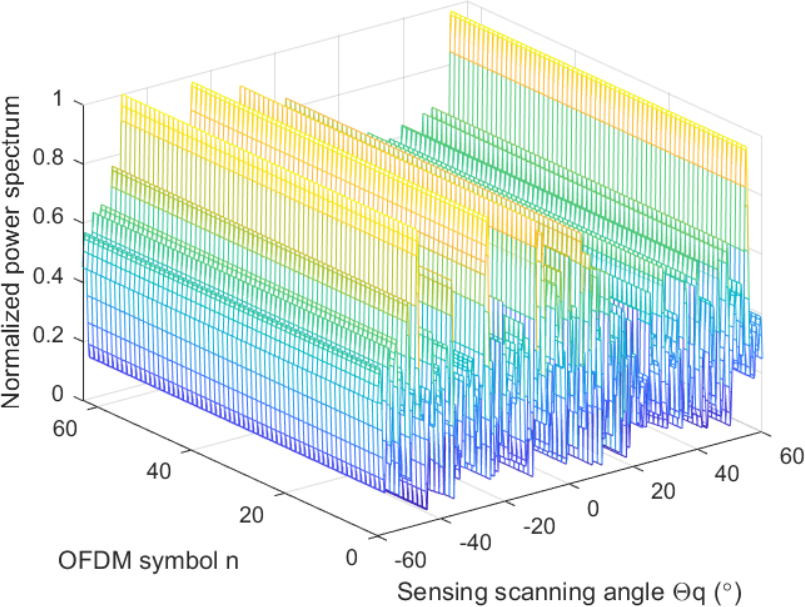}%
		\label{fig_first_case}}
	\hfil
	\subfloat[]{\includegraphics[width=70mm]{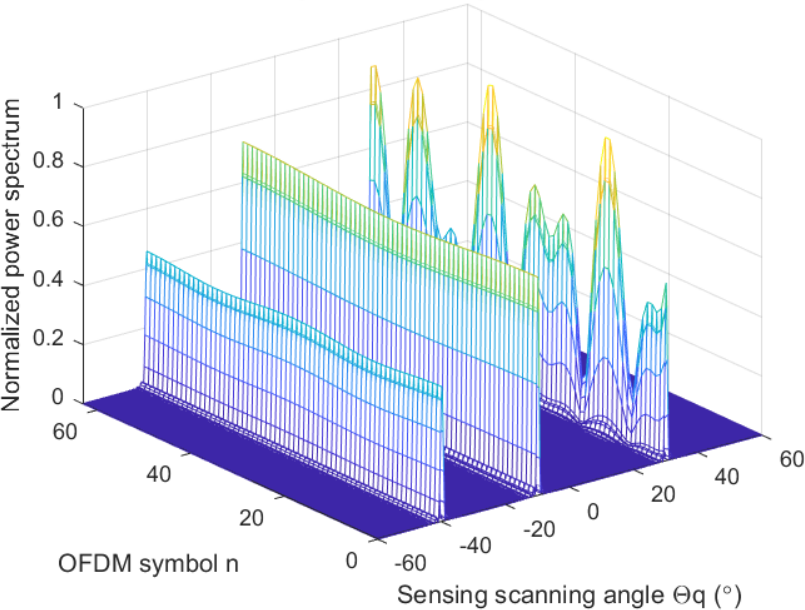}%
		\label{fig_first_case}}
	\hfil
	\subfloat[]{\includegraphics[width=70mm]{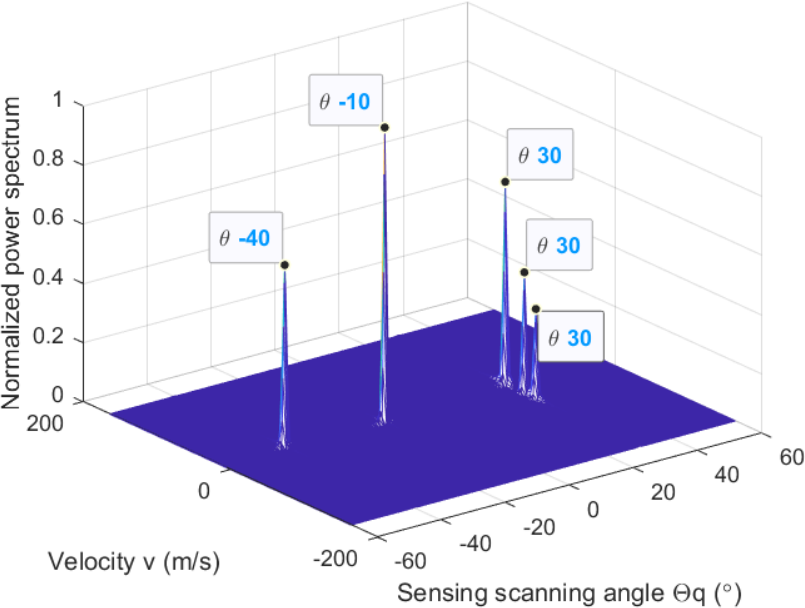}%
		\label{fig_first_case}}
	\hfil
	\subfloat[]{\includegraphics[width=70mm]{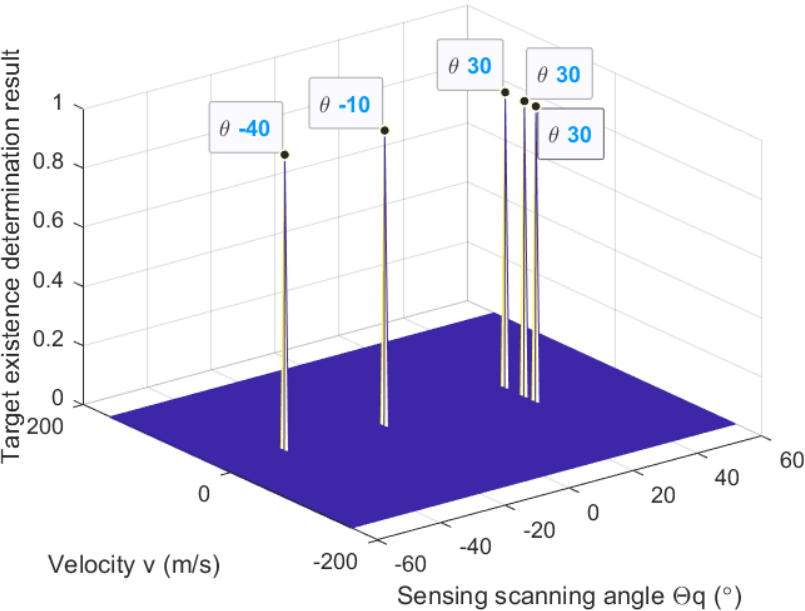}%
		\label{fig_second_case}}
	\caption{(a) The power spectrum of the original echo signals on a certain subcarrier. (b) Signal power spectrum after filtering out static environmental clutter.
		(c) The angle-Doppler spectrum.
		(d) The detection results of MSJD.
		The five dynamic targets are set as $(100m,-40^\circ,10m/s)$, $(60m,-10^\circ,5m/s)$, $(90m,30^\circ,15m/s)$, $(150m,30^\circ,-10m/s)$ and $(200m,30^\circ,-25m/s)$.}
	\label{fig_sim}
\end{figure*}

\subsection{Static Environmental Clutter Filtering}

It can be analyzed from (32) and (33) that
the echo signals ${\mathbf{Y}}_{cube}$ includes both DTE and SEE, 
and the corresponding EEC $\tilde{\mathbf{H}}_{cube}$  also includes both the
 EEC of  dynamic targets (DT-EEC) and the
 EEC of  static environment (SE-EEC).
As introduced in Section \RNum{1}, 
 the  sensing tasks of  ISAC system can be divided into  SES and  DTS,
while DTS can be considered as an estimation of DT-EEC.
However, when we focus on DTS problem, the SE-EEC in the original echo signals would cause negative interference to DTS, and thus  SE-EEC can be referred to as the clutter-EEC.
To address this negative interference,  we need  to filter out the interference of
clutter-EEC (SE-EEC) and  to extract the  effective DT-EEC from $\tilde{\mathbf{H}}_{cube}$.

To begin with,
the  EEC of the $m$-th subcarrier in $\tilde{\mathbf{H}}_{cube}$ can be denoted as  $\tilde{\mathbf{H}}_m^M = \tilde{\mathbf{H}}_{cube}[:,:,m] \in \mathbb{C}^{Q\times N}$, i.e.,
\begin{equation}
\begin{split}
\begin{aligned}
\label{deqn_ex1a}
&\tilde{\mathbf{H}}_m^M \! = \\& \! \!
\begin{bmatrix}\!\!
 \mathbf{w}_{R\!X,1}^H \mathbf{H}_{0,m} \tilde{\mathbf{x}}_{1,0,m}   & \!\cdots\!   & \mathbf{w}_{R\!X,1}^H \mathbf{H}_{N\!-\!1,m} \tilde{\mathbf{x}}_{1,N\!-\!1,m}   \\
 \!\vdots\!   & \!\ddots\!   & \vdots  \\
 \mathbf{w}_{R\!X,Q}^H \mathbf{H}_{0,m} \tilde{\mathbf{x}}_{Q,0,m}   &  \!\cdots\!    & \mathbf{w}_{R\!X,Q}^H \mathbf{H}_{N\!-\!1,m} \tilde{\mathbf{x}}_{Q,N\!-\!1,m} \\
\end{bmatrix}
\!\!+\! \mathbf{N}_m^M \!,
\end{aligned}
\end{split}
\end{equation}
where $\tilde{\mathbf{x}}_{q,n,m} = \mathbf{x}_{q,n,m}/s^{t,q}_{n,m}$, and the $(q,n)$-th element in $\mathbf{N}_m^M$ is $\mathbf{N}_m^M[q,n]=n^{s,q}_{n,m}/s^{t,q}_{n,m}$.
\textcolor{black}{Clearly, 
$\tilde{\mathbf{H}}_m^M$ includes both SE-EEC and DT-EEC. 
The idea is to first calculate the  SE-EEC as 
$\tilde{\mathbf{H}}^{static}_m$, 
and then subtract $\tilde{\mathbf{H}}^{static}_m$ from the total channel $\tilde{\mathbf{H}}_m^M$ to obtain the DT-EEC as $\tilde{\mathbf{H}}^{dynamic}_m\! =\! \tilde{\mathbf{H}}^M_m\! -\! \tilde{\mathbf{H}}^{static}_m$.}

Specifically, considering that the phase of  SE-EEC  remains  unchanged within multiple OFDM symbols, while the phase of  DT-EEC changes linearly with the target velocity within multiple OFDM symbols,
 we may average each row of   $\tilde{\mathbf{H}}_m^M$
and obtain the equivalent echoes channel vector caused by  static environment as $\tilde{\mathbf{h}}^{static}_m \in \mathbb{C}^{Q\times 1}$, \textcolor{black}{whose  $q$-th element is}
\begin{equation}
	\begin{split}\color{black}
		\begin{aligned}
			\label{deqn_ex1a}
&\!\!\tilde{\mathbf{h}}^{static}_m[q] \!=\! \tilde{h}^{static}_{m,q} \!=\! \frac{1}{N}\sum_{n=0}^{N-1} \tilde{h}^{s,q}_{n,m} \!=\!
\frac{1}{N}\sum_{n=0}^{N-1} {y}^{s,q}_{n,m}/s^{t,q}_{n,m} \\& 
\approx
\mathbf{w}_{RX,q}^H
\left(\sum_{i=1}^I \mathbf{H}_{i,n,m}'\right)\tilde{\mathbf{x}}_{q,n,m} + \tilde{n}^{static}_{m,q}.
		\end{aligned}
	\end{split}
\end{equation}
\textcolor{black}{The detailed derivation of  (36) can be found  in Appendix A.
Eq. (36) indicates that the average of $\tilde{h}_{n,m}^{s,q}$ for fixed $m$ and $q$  can be used to express the static clutter channel.}
Then the 
$\tilde{\mathbf{h}}^{static}_m$ can be used to reconstruct the SE-EEC as $\tilde{\mathbf{H}}^{static}_m=[\tilde{\mathbf{h}}^{static}_m,...,\tilde{\mathbf{h}}^{static}_m]\in \mathbb{C}^{Q\times N}$.
Then the DT-EEC of the $m$-th subcarrier can be extracted as
$\tilde{\mathbf{H}}^{dynamic}_m\! =\! \tilde{\mathbf{H}}^M_m\! -\! \tilde{\mathbf{H}}^{static}_m$, and we represent the
 $(q,n)$-th element in $\tilde{\mathbf{H}}^{dynamic}_m$ as 
\begin{equation}
\begin{split}
\begin{aligned}
\label{deqn_ex1a}
\!\!\! \check{h}_{q,n,m}\!=&
\mathbf{w}_{RX,q}^H\left(\sum_{k=1}^K\mathbf{H}_{k,n,m}\right)\tilde{\mathbf{x}}_{q,n,m}\!\\& \!\!\!-\!
\underbrace{\frac{1}{N}\!\!\sum_{n=0}^{N-1}\!\left[\mathbf{w}_{RX,q}^H\left(\sum_{k=1}^K\mathbf{H}_{k,n,m}\right)\tilde{\mathbf{x}}_{q,n,m}\right]}_{\mathcal{I}'_{q,n,m}} \!\!+\! \check{n}^{s,q}_{n,m},
\end{aligned}
\end{split}
\end{equation}
where 
$\mathcal{I}'_{q,n,m}$ is defined
as the remaining term after  clutter filtering, and $\check{n}^{s,q}_{n,m} = n^{s,q}_{n,m}/s^{t,q}_{n,m}-\frac{1}{N}\sum_{n=0}^{N-1}n^{s,q}_{n,m}/s^{t,q}_{n,m}$. \textcolor{black}{Because the Doppler term in DT-EEC varies linearly with continuous  symbols,
according to the derivation process in Appendix A, 
there is 
$\lim\limits_{N \to \infty} \mathcal{I}'_{q,n,m} =0$. 
Thus we have eliminated the influence of static environmental clutter.}

After filtering out  the static environmental clutter, the $\check{h}_{q,n,m}$ in (37) can be rewritten as (38) at  the bottom of this page, where $e^{j\phi^{k,q}_{n,m}}=e^{j\frac{4\pi f_0v_knT_s}{c}}e^{-j\frac{4\pi f_mr_k}{c}}$, and  $F_{RX}(\theta_1,\theta_2) \triangleq \mathbf{a}_{RX}^H(\theta_{1})\mathbf{a}_{RX}(\theta_2)$.
Considering the randomness of dynamic target  distribution,
we assume that there are $K_q$ dynamic targets simultaneously in the direction $\Theta_q$  with  $\sum_{q=1}^Q K_q = K$.
For massive MIMO systems with $N_T\to  \infty$ and $N_R\to  \infty$,
$\check{h}_{q,n,m}$ can be further calculated as (39) at  the bottom of this page.
Then the $\check{h}_{q,n,m}$ on all subcarriers of all OFDM symbols in all time slots can be stacked into a DT-EEC tensor $\tilde{\mathbf{H}}_{cube}^{dynamic}\in \mathbb{C}^{Q\times N\times M}$, 
whose $(q,n,m)$-th element is $\tilde{\mathbf{H}}_{cube}^{dynamic}[q,n,m] = \check{h}_{q,n,m}$.

\vspace{-3mm}

\subsection{Dynamic Targets Detection and Angle Estimation}

In order to detect the presence of dynamic targets from $\tilde{\mathbf{H}}_{cube}^{dynamic}$, we  extract the DT-EEC of the $m$-th subcarrier from $\tilde{\mathbf{H}}_{cube}^{dynamic}$ as
$\tilde{\mathbf{H}}_m^{dynamic} = \tilde{\mathbf{H}}_{cube}^{dynamic}[:,:,m]\in \mathbb{C}^{Q\times N}$.
Next, considering that the sensing scanning angle dimension in $\tilde{\mathbf{H}}_m^{dynamic}$ contains  angle information of dynamic targets while the OFDM symbol dimension contains  velocity information of dynamic targets,
we  perform an $N$-point fast fourier transform (FFT) on each row of $\tilde{\mathbf{H}}_m^{dynamic}$,
and  move the zero-frequency component of the transformed spectrum to the center of the array\cite{5776640,8443563}. By doing so, we can  obtain the angle-Doppler spectrum for  DT-EEC of the $m$-th subcarrier as
\setcounter{equation}{26}
\begin{equation}
\begin{split}
\begin{aligned}
\label{deqn_ex1a}
\tilde{\mathbf{H}}_m^{AD} = {\rm FFTshift}\{{\rm FFT}\{\tilde{\mathbf{H}}_m^{dynamic},N,2\},N,2\},
\end{aligned}
\end{split}\tag{40}
\end{equation}
\textcolor{black}{where ${\rm FFT}\{\tilde{\mathbf{H}}_m^{dynamic},N,2\}$ refers to performing $N$-point FFT on each row of $\tilde{\mathbf{H}}_m^{dynamic}$,
and ${\rm FFTshift}\{\mathbf{A},N,2\}$ refers to performing  $N$-point spectral centralization on each row of $\mathbf{A}$.}
For clarity,  we   introduce a corresponding matrix $\tilde{\mathbf{G}}_m^{AD}$ for $\tilde{\mathbf{H}}_m^{AD}$, which satisfies ${\rm \tilde{\mathbf{G}}}_m^{AD}[q,n] = |{\rm \tilde{\mathbf{H}}}_m^{AD}[q,n]|$.
 It can be analyzed from (39) that
when the resolutions of angle and velocity are sufficient, there will be $K$ peaks in $\tilde{\mathbf{G}}_m^{AD}$, which correspond to  $K$ dynamic targets one by one.

To further overcome the impact of noise and other factors, it is  necessary to set appropriate local thresholds to detect each  target from $\tilde{\mathbf{G}}_m^{AD}$, which is  referred to as \emph{radar target detection}.
Constant threshold (CT) detector  and constant false alarm rate (CFAR) detector are two main methods for  target detection\footnote{\textcolor{black}{The problem of radar target detection is to use the statistical characteristics of signals and  noise or clutter to solve the decision problem of whether the target signal is present or not.
CFAR detector aims to detect the targets in echo signals with a constant false alarm rate, that is, to accurately identify the presence of targets as much as possible while maintaining a low false alarm rate.
For this effect, 
CFAR  adaptively adjusting the detection threshold to maintain the false alarm rate, and it can adapt to different environments with different levels of background noise.}}\cite{4102829}. 
Here, we propose a joint detection method over multiple subcarriers (MSJD) based on CFAR and CT detectors,
which can be divided into the following two steps:
1) We  take  $\tilde{\mathbf{G}}_m^{AD}$ as the input, set appropriate reference  range and protection  range, and perform two-dimensional cell-averaging CFAR (2D-CA-CFAR) detection on $\tilde{\mathbf{G}}_m^{AD}$.
Then we can obtain the   judgment matrix over single subcarrier ${\rm\mathbf{G}}_m^{out}$ as the output.
2) We compute ${\rm\mathbf{G}}_m^{out}$ for each subcarrier, and  accumulate them together to obtain 
${\rm\mathbf{G}}^{out} = \sum_{m=1}^{M}{\rm\mathbf{G}}_m^{out}$.
Next we set a fixed threshold and perform CT detection on ${\rm\mathbf{G}}^{out}$ to obtain the   judgment matrix over multiple subcarriers as ${\rm\mathbf{G}}^{out}_{mask}$.
Then we can detect each target from ${\rm\mathbf{G}}^{out}_{mask}$,
and  the accurate angle estimation results  for each dynamic target can also be obtained from ${\rm\mathbf{G}}^{out}_{mask}$
as  $\{\hat{\theta}_1,\hat{\theta}_2,...,\hat{\theta}_K\}$.
\textcolor{black}{For clarity, Fig.~7 shows a simple application example of the proposed MSJD algorithm.}

Fig.~8 shows  examples of   clutter filtering,
angle-Doppler spectrum estimation (ADSE) and MSJD under noiseless conditions, where  five dynamic   targets are set as
 $(100m,-40^\circ,10m/s)$, $(60m,-10^\circ,5m/s)$, $(90m,30^\circ,15m/s)$, $(150m,30^\circ,-10m/s)$ and $(200m,30^\circ,-25m/s)$.
It is seen from Fig.~8(a) that the original echo signals  carry dense clutter caused by  static environment. After clutter filtering, Fig.~8(b)  only retains the effective echoes caused by  dynamic targets.
After performing velocity FFT on the complex signals, five peaks corresponding to five dynamic targets can be clearly observed in the angle-Doppler spectrum shown in Fig.~8(c), which can also be observed in the MSJD  results shown in Fig.~8(d).
 Then the angle estimates of the dynamic targets  can be obtained as $-40^\circ$, $-10^\circ$, $30^\circ$, $30^\circ$ and $30^\circ$, respectively.

\subsection{Distance and Velocity Estimation for Dynamic Targets}

\begin{figure*}[!t]
\centering
\subfloat[]{\includegraphics[width=75mm]{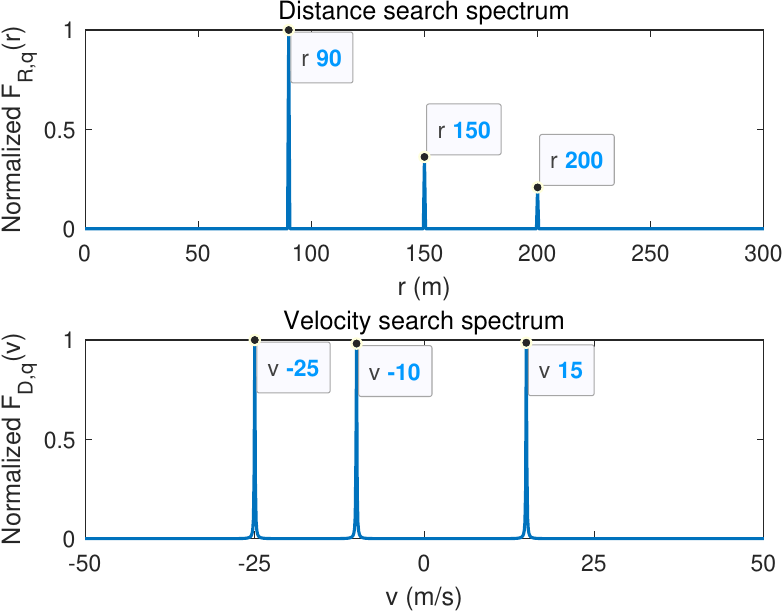}%
\label{fig_first_case}}
\hfil
\subfloat[]{\includegraphics[width=96mm]{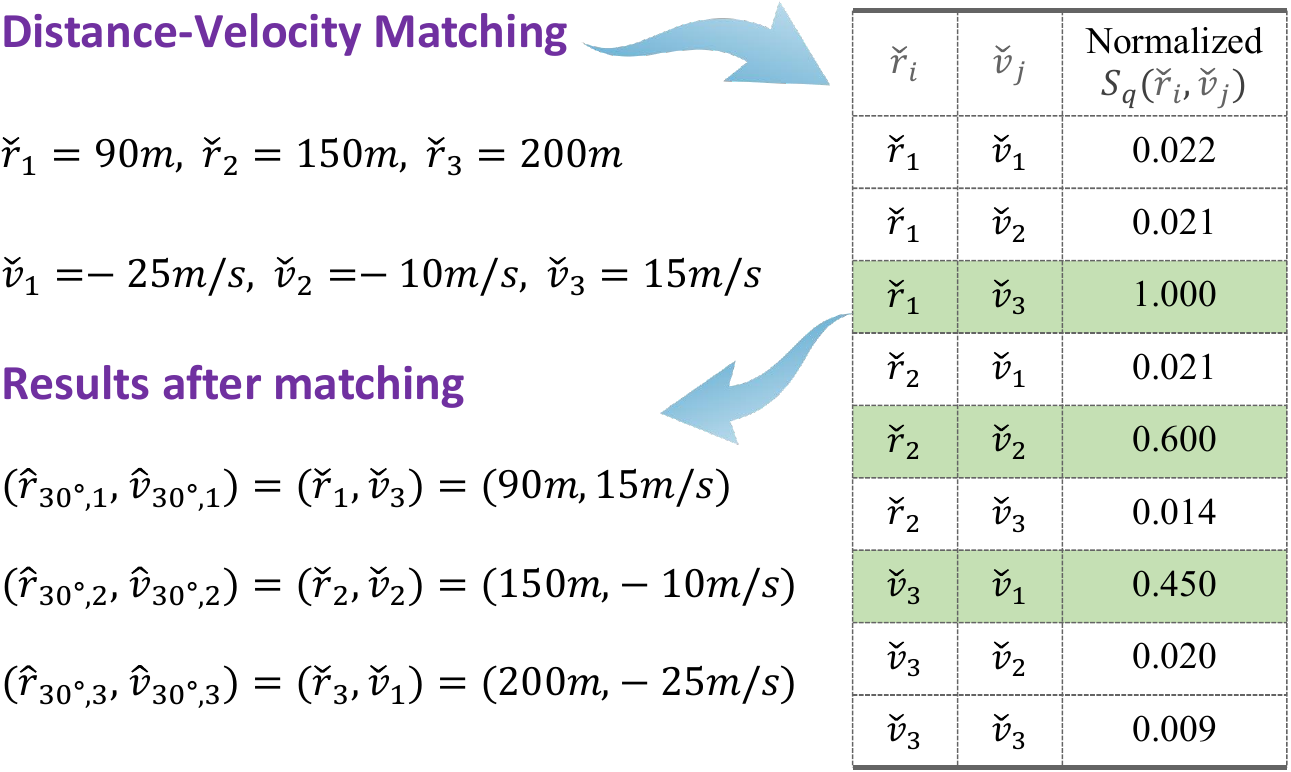}%
\label{fig_second_case}}
\caption{(a) An example of  velocity estimation based on searching $F_{D,q}(v)$;
(b) An example of  distance estimation based on searching $F_{R,q}(r)$.
Three dynamic targets are set as $(90m,30^\circ,15m/s)$, $(150m,30^\circ,-10m/s)$ and $(200m,30^\circ,-25m/s)$.}
\label{fig_sim}
\end{figure*}

After obtaining  angle estimates, we may suppose that $K_q$ detected  targets are located in the same angle $\Theta_q$,
and  represent the parameters of the $k'_q$-th dynamic target as $(r_{k_q'},\Theta_q,v_{k_q'})$, where $k'_q = 1,2,...,K_q$.
Then the DT-EEC of this beam scanning angle within all subcarriers and  symbols can be represented as $\tilde{\mathbf{H}}^{RD}_q \in \mathbb{C}^{N\times M}$, whose  $(n,m)$-th element is $\tilde{\mathbf{H}}^{RD}_q[n,m]=\check{h}_{q,n,m}$.
Based on  (39), $\tilde{\mathbf{H}}^{RD}_q$ and its transpose matrix  $(\tilde{\mathbf{H}}^{RD}_q)^T$  can be respectively represented as
\begin{equation}
\begin{split}
\begin{aligned}
\label{deqn_ex1a}
\tilde{\mathbf{H}}^{RD}_q\! =\zeta_q \sum_{k'_q=1}^{K_q}
\alpha'_{k'_q} \mathbf{k}_{D}(v_{k_q'}) \mathbf{k}^T_{R}(r_{k_q'})+ \mathbf{N}_q^{RD},
\end{aligned}
\end{split}\tag{41}
\end{equation}
\begin{equation}
\begin{split}
\begin{aligned}
\label{deqn_ex1a}
(\tilde{\mathbf{H}}^{RD}_q)^T\! =\zeta_q \sum_{k'_q=1}^{K_q}
\alpha'_{k'_q} \mathbf{k}_{R}(r_{k_q'}) \mathbf{k}^T_{D}(v_{k_q'}) + (\mathbf{N}_q^{RD})^T,
\end{aligned}
\end{split}\tag{42}
\end{equation}
where $\zeta_q$ is the amplitude term, $\alpha'_{k'_q} = 
\alpha_{k'_q}e^{-j\frac{4\pi f_0r_{k'_q}}{c}}$,
$\mathbf{N}_q^{RD}$  is the corresponding noise matrix,
$\mathbf{k}_{D}(v)=[1,e^{j\frac{4\pi f_0vT_s}{c}},...,e^{j\frac{4\pi f_0vT_s}{c}(N-1)}]^T \in \mathbb{C}^{N\times 1} $ is defined as the \emph{Doppler array steering vector}, and $\mathbf{k}_{R}(r)=[1,e^{-j\frac{4\pi r\Delta f}{c}},...,e^{-j\frac{4\pi r\Delta f}{c}(M-1)}]^T \in \mathbb{C}^{M\times 1} $ is defined as the \emph{distance array steering vector}.
From  (40) and (41), it can be seen that $\tilde{\mathbf{H}}^{RD}_q$ and $(\tilde{\mathbf{H}}^{RD}_q)^T$ are the generalized array signals  related to Doppler array and distance array, respectively. 
To estimate the velocity and  distance of dynamic targets, the autocorrelation matrices of $\tilde{\mathbf{H}}^{RD}_q$ and $(\tilde{\mathbf{H}}^{RD}_q)^T$ are calculated as $\mathbf{R}^X_{D,q} =\frac{1}{M}\tilde{\mathbf{H}}^{RD}_q  (\tilde{\mathbf{H}}^{RD}_q )^H$ and $\mathbf{R}^X_{R,q} =\frac{1}{N}(\tilde{\mathbf{H}}^{RD}_q)^T  ((\tilde{\mathbf{H}}^{RD}_q )^T)^H$. We perform eigenvalue decomposition of  $\mathbf{R}^X_{D,q}$ and $\mathbf{R}^X_{R,q}$ to obtain the diagonal matrix with eigenvalues ranging from large to small ($\mathbf{\Sigma}_{D,q}$ and $\mathbf{\Sigma}_{R,q}$) and the corresponding eigenvector matrix ($\mathbf{U}_{D,q}$ and $\mathbf{U}_{R,q}$). That is $[\mathbf{U}_{D,q}, \mathbf{\Sigma}_{D,q}]={\rm eig}(\mathbf{R}^X_{D,q})$ and $[\mathbf{U}_{R,q}, \mathbf{\Sigma}_{R,q}]={\rm eig}(\mathbf{R}^X_{R,q})$.
Then  the minimum description length (MDL) criterion is utilized to estimate the number of dynamic targets from $\mathbf{\Sigma}_{D,q}$ and $\mathbf{\Sigma}_{R,q}$  as $K_{q,D}^{MDL}$ and $K_{q,R}^{MDL}$ respectively\cite{mdl,mdl2}, and  we define $K_{q}^{MDL}={\rm min}\{K_{q,D}^{MDL}, K_{q,R}^{MDL}\}$ as the number of dynamic targets to be estimated from $\mathbf{{H}}_{q}^{RD}$.
Therefore, the noise space related to the Doppler array can be represented as $\mathbf{U}^N_{D,q} = \mathbf{U}_{D,q}[:,K_{q}^{MDL}+1:N]$, and the noise space related to the distance array can be represented as $\mathbf{U}^N_{R,q} = \mathbf{U}_{R,q}[:,K_{q}^{MDL}+1:M]$.
Then the  Doppler spectral function with search velocity $v$  and the  distance spectral function with search distance $r$ can be defined as
\begin{equation}
\begin{split}
\begin{aligned}
\label{deqn_ex1a}
F_{D,q}(v) = \frac{1}{\mathbf{k}^H_{D}(v)\mathbf{U}^N_{D,q}(\mathbf{U}^N_{D,q})^H\mathbf{k}_{D}(v)},
\end{aligned}
\end{split}\tag{43}
\end{equation}
\begin{equation}
\begin{split}
\begin{aligned}
\label{deqn_ex1a}
F_{R,q}(r) = \frac{1}{\mathbf{k}^H_{R}(r)\mathbf{U}^N_{R,q}(\mathbf{U}^N_{R,q})^H\mathbf{k}_{R}(r)}.
\end{aligned}
\end{split}\tag{44}
\end{equation}
By searching for the peaks of  $F_{D,q}(v)$ and $F_{R,q}(r)$, we can obtain the preliminary velocity estimates for $K_q$ targets as $\{\check{v}_1,\check{v}_2,...,\check{v}_{K_q}\}$ and the preliminary distance estimates for $K_q$ targets as $\{\check{r}_1,\check{r}_2,...,\check{r}_{K_q}\}$, respectively. 
Next, in order to obtain the  unique parameters information for each target from the preliminary estimates,
we must realize   distance and velocity matching from  $\{\check{v}_1,\check{v}_2,...,\check{v}_{K_q}\}$  and  $\{\check{r}_1,\check{r}_2,...,\check{r}_{K_q}\}$. Specifically, we construct a distance-velocity set as $\Xi_q = \{
(\check{r}_1,\check{v}_1),...,(\check{r}_1,\check{v}_{K_q}),...,(\check{r}_{K_q},\check{v}_1),...,(\check{r}_{K_q},\check{v}_{K_q})\}$ containing $K_q^2$ elements.
Then the matching value of the element $(r_i,v_j)$ in $\Xi_q$ can be calculated as
\begin{equation}
\begin{split}
\begin{aligned}
\label{deqn_ex1a}
S_q(r_i,v_j) = |{\mathbf{k}^H_{D}(v_j)\mathbf{H}^{RD}_{q}\mathbf{k}^*_{R}(r_i)}|.
\end{aligned}
\end{split}\tag{45}
\end{equation}
After performing (45) for all $r_i$ and $v_j$ pairs,  the set of matching values corresponding to the set $\Xi_q$ can be represented as  $\Xi_q^{S}$. 
By searching for the $K_q$ elements with the highest value in $\Xi_q^{S}$ and their corresponding elements in $\Xi_q$, we can estimate the  distances and velocities of these $K_q$ dynamic targets as $\{(\hat{r}_{q,1},\hat{v}_{q,1}),(\hat{r}_{q,2},\hat{v}_{q,2}),...,(\hat{r}_{q,K_q},\hat{v}_{q,K_q})\}$.

Fig.~9 shows  examples of  distance and velocity estimation and distance-velocity matching under noiseless conditions,  for  estimation of  three targets located in  $30^\circ$  direction set in Fig.~8.
The preliminary distance can be estimated  from the distance search spectrum in Fig.~9(a) as $90m$, $150m$ and $200m$, while the 
preliminary velocity can be estimated  from the velocity search spectrum
as $-25m/s$, $-10m/s$ and $15m/s$.
Fig.~9(b) shows the process of distance-velocity matching using  (45), and we can  obtain the  estimation results for the three dynamic targets located in  $30^\circ$  direction, which are $(90m,30^\circ,15m/s)$, $(150m,30^\circ,-10m/s)$ and $(200m,30^\circ,-25m/s)$.

\section{Simulation Results}

In simulations, we set 
the number of antennas for  transmitting array as $N_T=128$, 
the number of antennas for  receiving array as $N_R=128$,
the lowest carrier frequency as $f_0 = 220$ GHz and the antenna spacing as $d=\frac{1}{2}\lambda$.
The noise is assumed to obey the complex Gaussian distribution with mean $\mu=0$ and  variance $\sigma_n^2=1$.
The root  mean square error (RMSE) of angle estimation,  distance estimation,  and velocity estimation are defined as
${\rm RMSE}_\theta=\sqrt{\frac{\sum _{i=1}^{Count}(\hat{\theta}_{s(i)}-\theta_{s})^2}{Count}}$, ${\rm RMSE}_r=\sqrt{\frac{\sum _{i=1}^{Count}(\hat{r}_{s(i)}-r_{s})^2}{Count}}$ and ${\rm RMSE}_v=\sqrt{\frac{\sum _{i=1}^{Count}(\hat{v}_{s(i)}-v_{s})^2}{Count}}$,
where $Count$ is the number of the Monte Carlo runs,
the real parameters of the dynamic target is
 $(r_{s},\theta_{s},v_s)$, and $(\hat{r}_{s(i)},\hat{\theta}_{s(i)},\hat{v}_{s(i)})$ is the estimation parameters of the target.
 
 \textcolor{black}{The required sensing range of ISAC BS is set as 
 $\{(r,\theta)|r_{min}\leq r \leq r_{max},-60^\circ \leq \theta \leq 60^\circ\}$, in which the range of distance sensing varies slightly in different scenes.
 	We assume that there is clutter everywhere within the sensing range of the BS.
 	The size of the clutter scattering unit is determined by the system's sensing resolution. Specifically, the angle resolution is approximately $\Delta \theta \approx \frac{2}{N_T}$, and the distance resolution is $\Delta r = \frac{c}{2(M-1)\Delta f}$.}
 	It is worth pointing out that existing ISAC schemes \cite{9898900,10050406,8918315,8114253,10048770} that do not consider clutter environment cannot detect the targets and estimate their parameters in actual clutter environment as shown in Fig.~8(a).
 	Thus, we did not show the performance of the existing methods in the simulations.

\begin{figure}[!t]
\centering
\includegraphics[width=80mm]{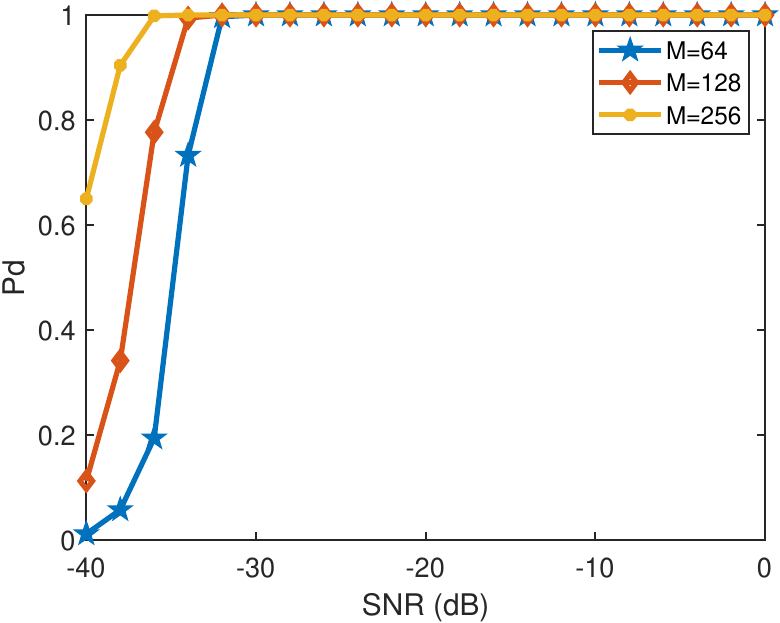}
\caption{Curve of dynamic target detection probability versus SNR.}
\label{fig_1}
\end{figure}

\begin{figure*}[!t]
\centering
\subfloat[]{\includegraphics[width=60mm]{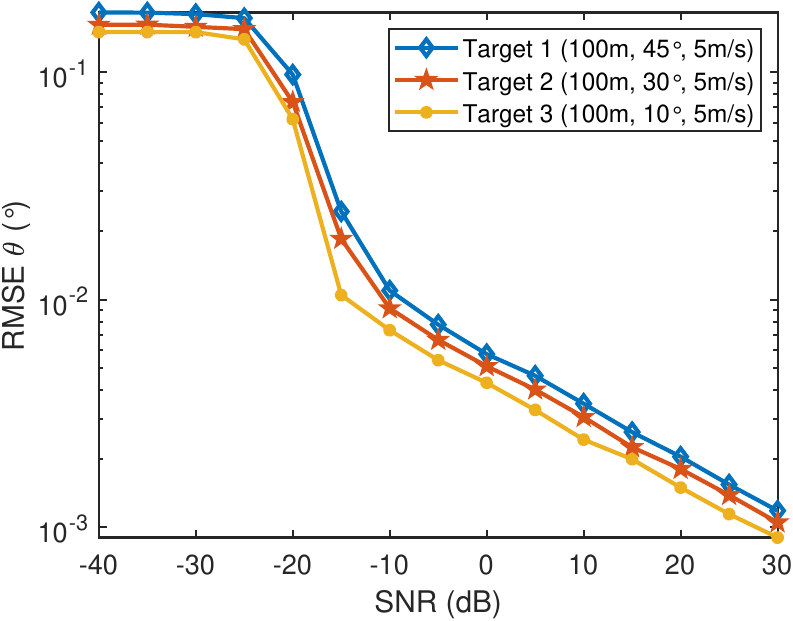}%
\label{fig_first_case}}
\hfil
\subfloat[]{\includegraphics[width=60mm]{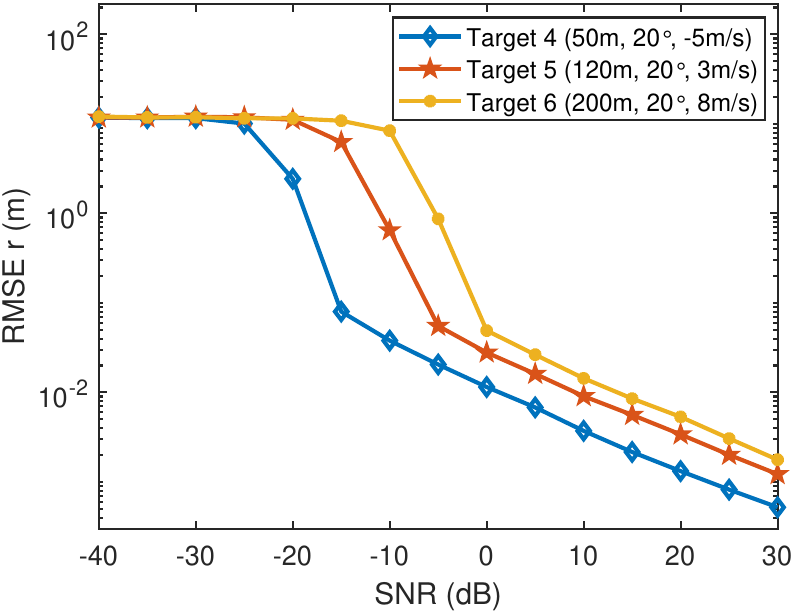}%
\label{fig_second_case}}
\hfil
\subfloat[]{\includegraphics[width=60mm]{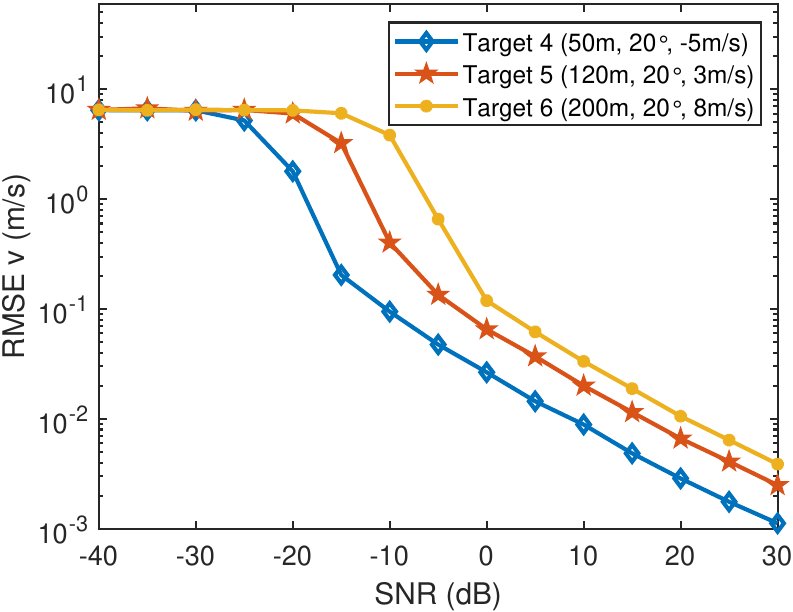}%
\label{fig_second_case}}
\caption{(a) The RMSE of dynamic target angle sensing;
(b) The RMSE of dynamic target distance sensing; (c) The RMSE of dynamic target velocity sensing.  \textcolor{black}{The parameter settings for dynamic targets are shown in the figures.}}
\label{fig_sim}
\end{figure*}

\subsection{Performance of  Dynamic Target Detection}
We set that the subcarrier frequency interval is
$\Delta f = 200$ kHz, and the number of OFDM symbols is $N=64$. 
\textcolor{black}{To test the performance of dynamic target  detection, we  generate several dynamic targets in each repeated experiment, which are randomly and uniformly distributed within the sensing range of the BS. }
Fig.~10 shows the performance curve of
dynamic target detection probability $P_d$
  over multiple subcarriers  with different subcarrier numbers.
It can be seen that   $P_d$ gradually increases with the increase of SNR. 
When the SNR is greater than a certain value, $P_d$ can almost approach $100\%$.
In addition, when the SNR is low, $P_d$ increases with the increase of  $M$, which verifies the effectiveness of the proposed joint detection method over multiple subcarriers.

\subsection{Performance of Dynamic Target Parameter Estimation}

We assume that the subcarrier frequency interval is
$\Delta f = 500$ kHz, the number of OFDM symbols is
$N=64$, and the number of subcarriers is $M=128$.
Fig.~11 shows the curves of sensing RMSE versus SNR for  different dynamic targets.

It can be seen from Fig.~11(a) that the ${\rm RMSE}_\theta$ gradually decreases with the increase of SNR.
When the SNR is less than $-25$ dB, the ${\rm RMSE}_\theta$ is large and there is no significant change for ${\rm RMSE}_\theta$ with the change of SNR.  As the SNR increases from $-25$ dB to $-10$ dB, the ${\rm RMSE}_\theta$ rapidly decreases from $0.15^\circ$ to $0.01^\circ$. As the SNR further increases, the ${\rm RMSE}_\theta$ continues to decrease.
In addition, when the distance and velocity parameters are fixed, the closer the angle of the dynamic target approaches $0^\circ$, the smaller its ${\rm RMSE}_\theta$ will be. This is mainly because the beam generated by antenna array is narrower near $0^\circ$, which improves the performance of angle sensing to some extent.

Fig.~11(b) shows the variation curve of ${\rm RMSE}_r$ versus SNR, where three dynamic targets are fixed in  $20^\circ$ direction, but their distances are different. 
It can be seen that when the SNR is low, 
the ${\rm RMSE}_r$  remains at a large value, which means
that it is difficult to estimate the target parameters under harsh SNR conditions. 
Nevertheless, when the SNR is greater than $0$ dB, the proposed sensing scheme can accurately estimate the distance of the three targets.

Fig.~11(c) shows the RMSE variation curve of target velocity sensing versus SNR.
For the target at $50m$, when the SNR reaches $-10$ dB, its velocity sensing RMSE is about $0.1m/s$. 
When the SNR increases to $10$ dB, the ${\rm RMSE}_v$  decreases to $0.01m/s$, which indicates that the proposed sensing scheme has high velocity sensing accuracy.

By comparing the sensing results of three targets at different distances in Fig.~11(b) and Fig.~11(c), we can find that the sensing error of targets at farther distances is larger. This is mainly because the signal power of the dynamic target echo decreases with the increase of target distance, and thus, farther targets will be  more susceptible to noise.

\subsection{The Impact of System Resolution on Sensing Performance}

The sensing accuracy of a system is usually related to its sensing resolution. Here we will explore the impact of system parameters on sensing resolution and performance.

\begin{figure*}[!t]
\centering
\subfloat[]{\includegraphics[width=60mm]{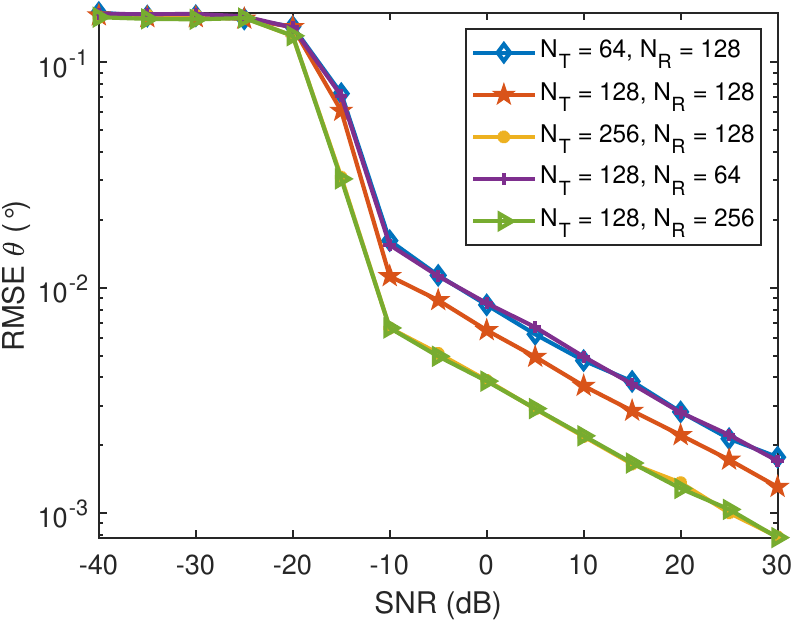}%
\label{fig_first_case}}
\hfil
\subfloat[]{\includegraphics[width=60mm]{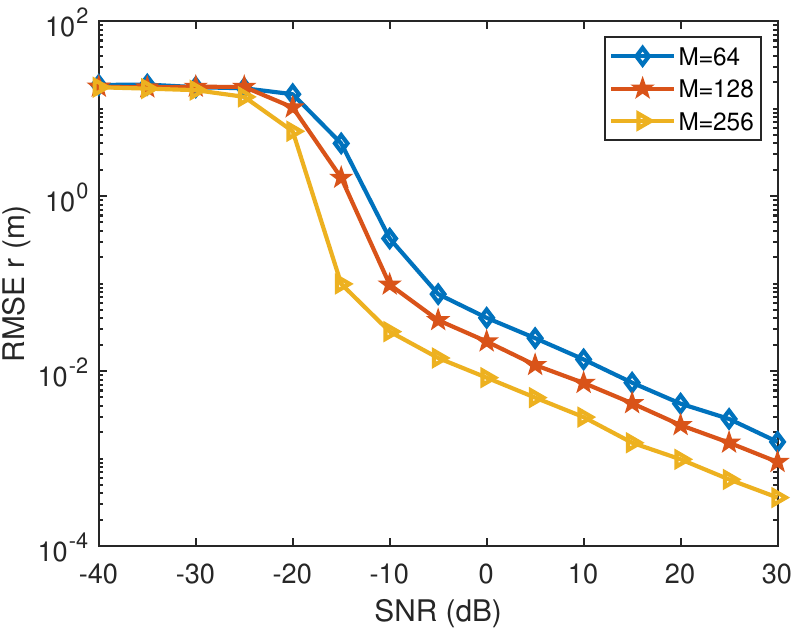}%
\label{fig_second_case}}
\hfil
\subfloat[]{\includegraphics[width=60mm]{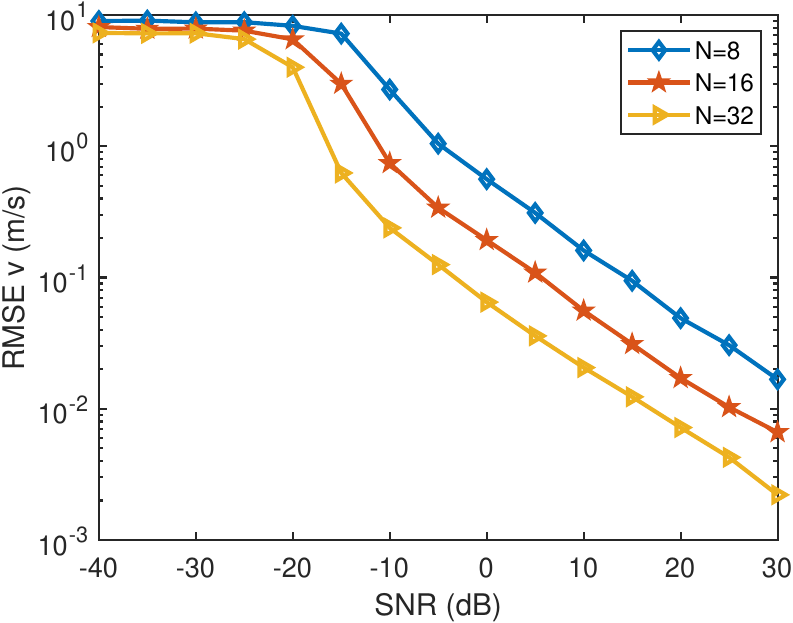}%
\label{fig_second_case}}
\caption{(a) The RMSE for angle sensing under different array sizes;
(b) The RMSE for distance sensing under different
numbers of subcarriers; (c) The RMSE  for velocity sensing under different numbers of OFDM symbols. \textcolor{black}{The dynamic target is set as $(110m, -20^\circ, 10m/s)$.}}
\label{fig_sim}
\end{figure*}

 Fig.~12(a) shows the variation of RMSE versus SNR for  angle sensing under different array sizes,
in which  the subcarrier frequency interval is $\Delta f=500$ kHz, the number of  symbols is $N=32$ and the number of subcarriers is $M=128$.
 It is seen  that when the number of transmitting array antennas $N_T$ or the number of receiving array antennas $N_R$ is fixed, the accuracy of angle sensing gradually improves with the increase of $N_R$ or $N_T$.
This phenomenon is mainly because the angle resolution of MIMO systems is  inversely proportional to the number of antennas, and more antennas can form narrower beams to provide higher sensing performance.

Fig.~12(b) shows the variation of RMSE versus SNR for  distance sensing under different 
numbers of subcarriers,
in which  the array size are $N_T=128$ and $N_R=128$,
 subcarrier frequency interval is $\Delta f=500$ kHz,
and the number of OFDM symbols is $N=32$.
It can be seen from the figure that the RMSE of distance sensing significantly decreases with the increase of the number of subcarriers $M$.
In fact, when $\Delta f$ is fixed, the system transmission bandwidth $W=(M-1)\Delta f$ linearly increases with $M$, and the  distance resolution $\Delta r = \frac{c}{2B}=\frac{c}{2(M-1)\Delta f}$ gradually decreases in numerical value\cite{10011222}. Hence,  more subcarriers can realize higher distance resolution, which in turn leads to higher distance sensing accuracy.

Fig.~12(c) shows the variation of RMSE versus SNR for  velocity sensing under different 
numbers of OFDM symbols,
in which  the array size are $N_T=128$ and $N_R=128$,
 subcarrier frequency interval is $\Delta f=500$ kHz,
and the number of subcarriers is $M=128$.
It can be observed that when  $N$ is not very large, the accuracy of  velocity sensing is already quite satisfactory. In addition,
 the RMSE of velocity sensing  decreases significantly with the increase of  $N$.
In reality,  when $\Delta f$ is fixed,
the velocity resolution of the system $\Delta v = \frac{c}{2f_0NT_s}$  decreases  as $N$ increases\cite{10011222}.
Hence, more OFDM symbols can bring higher velocity resolution and achieve higher accuracy in velocity sensing.

\subsection{Communications Performance Evaluation}

In this subsection, we will evaluate the communications performance of the proposed ISAC process by using the bit error rate (BER)  as the indicator.
We  set  three communication users located in the 
$-40^\circ$, $0^\circ$, and $30^\circ$ directions.
One sensing beam scans the entire space, and 
three communications beams communicate with the users for data  transmission,  in which  16-QAM is employed for communications signals modulation. The number of subcarriers is set to $M=512$,
the subcarrier frequency interval is set to $\Delta f=480$ kHz,
 the number of OFDM symbols is set to $N=64$,
and the number of BS transmitting antennas $N_T$
 is set as a variable.

\begin{figure}[!t]
\centering
\includegraphics[width=80mm]{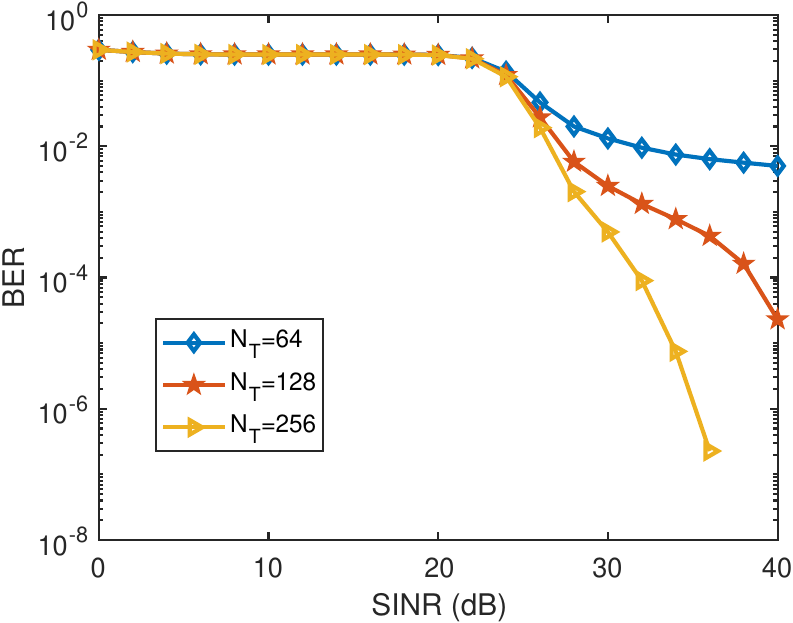}
\caption{The curve of BER versus SINR under different array sizes.}
\label{fig_1}
\end{figure}

Fig.~13 shows the curve of BER versus SINR under different array sizes. 
It can be seen that when the SINR of the  user is high enough, BER gradually decreases as the SINR increases. However, when $N_T$ is small, BER will converge to an error floor as SINR increases. This is mainly because when the antenna array is small, it is difficult to alleviate inter users interference and sensing interference to communications.
In addition,  it can be found that BER gradually decreases as $N_T$ increases. When $N_T$ is $256$ and SINR is $40$ dB, the overall BER of communication in the simulation is almost zero.

\begin{figure}[!t]
\centering
\subfloat[]{\includegraphics[width=80mm]{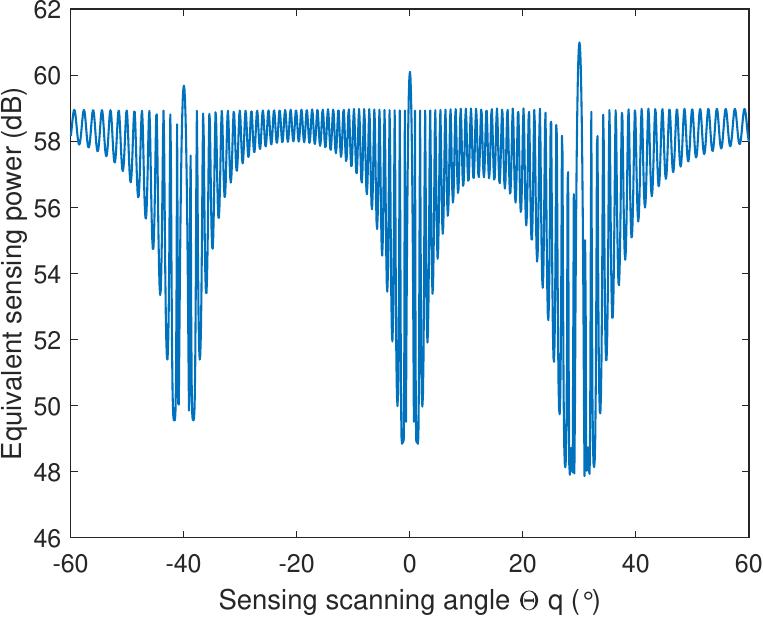}%
\label{fig_first_case}}
\vfil
\subfloat[]{\includegraphics[width=80mm]{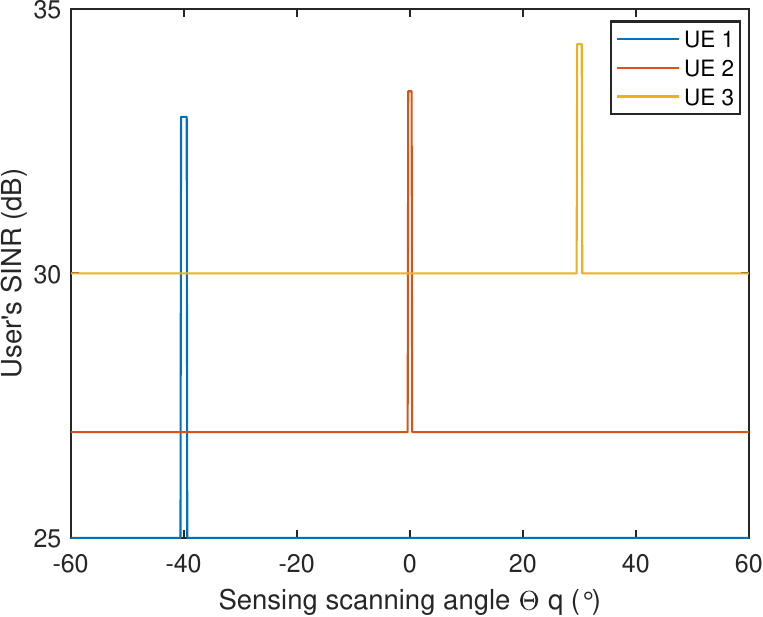}%
\label{fig_second_case}}
\caption{\textcolor{black}{(a) The variation of equivalent sensing power during SBS  process;
(b) The variation of  user's SINR during SBS  process.  Three users are set up in the simulation, and their minimum SINR requirements are $25$ dB, $27$ dB, and $30$ dB, respectively.}}
\label{fig_sim}
\end{figure}

\subsection{\textcolor{black}{Stability of Sensing Beam Scanning Process}}

\textcolor{black}{
We set the system parameters to $N_T = 128$, $N_R=128$, $f_0=100$ GHz, $P_t = 100$. Besides, we set three users located at $(60m, -40^\circ)$, $(60m, 0^\circ)$ and $(60m, 30^\circ)$, and their minimum SINR requirements are $25$ dB, $27$ dB, and $30$ dB, respectively.
Then Fig.~14 shows the performance variation curve of sensing and communications during the sensing beam scanning process.
It can be intuitively seen from the figure that the proposed  power allocation algorithm can effectively meet the SINR requirements of communications users, while providing satisfactory equivalent sensing  power.}

\section{Conclusions}

In this paper, we have proposed a practical ISAC framework to sense    dynamic targets from   clutter environment while ensuring users communications quality.
In order to implement  communications function and sensing function simultaneously,  we have designed multiple communications beams that can communicate with the users and one rotating sensing beam can scan the entire  space. 
To minimize the interference of sensing beam on existing communications systems, we have divided the service area into S4S sectors and  C4S sectors, and have provided beamforming design and power allocation optimization strategies for each type sector.
Then we have proposed the clutter model for  ISAC system in urban scenarios and  have derived the sensing  channel model that includes both  static environment and  dynamic targets.
When  BS received  echo signals,
we first 
filter out the  static environmental clutter and extract the effective dynamic target echoes.
Then the task of dynamic target sensing  is to detect the presence of dynamic targets and to estimate their angles, distances, and velocities parameters.
Among them,
dynamic target detection and angle estimation were realized through
 angle-Doppler spectrum estimation and joint detection  over multiple subcarriers, while distance and velocity estimation were realized through the extended subspace algorithm.
Simulation results have been provided to demonstrate the effectiveness of the proposed schemes and its superiority over the existing methods that ignore 
clutter environment.

\section*{\textcolor{black}{APPENDIX A \\ Proof of (36)}}

\textcolor{black}{
To demonstrate the calculation process more clearly, let us ignore the impact of noise, and then  $\tilde{h}^{s,q}_{n,m}$ is represented as}
\begin{equation}
	\begin{split}\color{black}
		\begin{aligned}
			\label{deqn_ex1a}
&\tilde{h}^{s,q}_{n,m} = y^{s,q}_{n,m}/s^{t,q}_{n,m} \!=\! \mathbf{w}_{RX,q}^H \mathbf{H}_{n,m} \tilde{\mathbf{x}}_{q,n,m}\\&=
\mathbf{w}_{RX,q}^H
\left(\sum_{k=1}^K \mathbf{H}_{k,n,m} + \sum_{i=1}^I \mathbf{H}_{i,n,m}'\right)\tilde{\mathbf{x}}_{q,n,m} \\&=
\sum_{k=1}^K  \mathbf{w}_{RX,q}^H \mathbf{H}_{k,n,m} \tilde{\mathbf{x}}_{q,n,m} + \sum_{i=1}^I \mathbf{w}_{RX,q}^H  \mathbf{H}_{i,n,m}' \tilde{\mathbf{x}}_{q,n,m} \\& = 
\sum_{k=1}^K \alpha_k e^{j \!\frac{4\pi f_0v_k}{c}\!nT_s} e^{-j\!\frac{4\pi f_mr_k}{c}}
\mathbf{w}_{RX,q}^H \mathbf{a}_{R\!X}(\theta_k)\mathbf{a}_{T\!X}^H(\theta_k)\tilde{\mathbf{x}}_{q,n,m} \\ & \quad \quad + \sum_{i=1}^I  \beta_i e^{-j2\pi f_m\!\frac{2r_i'}{c}} \mathbf{w}_{RX,q}^H \mathbf{a}_{R\!X}(\theta_i')\mathbf{a}_{T\!X}^H(\theta_i') \tilde{\mathbf{x}}_{q,n,m}.
		\end{aligned}
	\end{split}\tag{46}
\end{equation}
\textcolor{black}{Note that 
$\mathbf{w}_{R\!X,q}^H \mathbf{a}_{R\!X}(\theta_k)\mathbf{a}_{T\!X}^H(\theta_k)\tilde{\mathbf{x}}_{q,n,m}$ and 
$\mathbf{w}_{RX,q}^H \mathbf{a}_{R\!X}(\theta_i')\mathbf{a}_{T\!X}^H(\theta_i') \tilde{\mathbf{x}}_{q,n,m}$
have eliminated the influence of random symbols
 under  massive MIMO array,   and thus $\mathbf{w}_{RX,q}^H \mathbf{a}_{R\!X}(\theta_k)\mathbf{a}_{T\!X}^H(\theta_k)\tilde{\mathbf{x}}_{q,n,m}$ and 
 $\mathbf{w}_{RX,q}^H \mathbf{a}_{R\!X}(\theta_i')\mathbf{a}_{T\!X}^H(\theta_i') \tilde{\mathbf{x}}_{q,n,m}$ are  independent of $n$.
Then the average of $\tilde{h}^{s,q}_{n,m}$ for the fixed $m$ and $q$ can be calculated as}
\begin{equation}
	\begin{split}\color{black}
		\begin{aligned}
			\label{deqn_ex1a}
& \tilde{h}^{static}_{m,q} = \tilde{\mathbf{h}}^{static}_m[q] = \frac{1}{N}\sum_{n=0}^{N-1} \tilde{h}^{s,q}_{n,m} 
\\&=\!\!\frac{1}{N}\!\!\!\sum_{n=0}^{N-1}\!\!\left[ \!\sum_{k=1}^K \!\alpha_k e^{j \!\frac{4\pi\! f_0\!v_k}{c}\!nT_s}\! e^{\!-\!j\!\frac{4\pi \!f_m\!r_k}{c}}
\!\mathbf{w}_{\!R\!X\!,q}^H \mathbf{a}_{R\!X}(\!\theta_k\!)\mathbf{a}_{T\!X}^H(\!\theta_k\!)\tilde{\mathbf{x}}_{q\!,n\!,m} \!\!\right] \\ & \quad \quad +\!\frac{1}{N}\!\!\sum_{n=0}^{N-1}\!\!\left[ \sum_{i=1}^I  \!\beta_i e^{-j2\pi f_m\!\frac{2r_i'}{c}} \mathbf{w}_{R\!X,q}^H \mathbf{a}_{R\!X}(\theta_i')\mathbf{a}_{T\!X}^H(\theta_i') \tilde{\mathbf{x}}_{q\!,n\!,m}\! \right]\\&=\!\!\frac{1}{N}\!\!\!\sum_{n=0}^{N-1}\!\!\left[ \!\sum_{k=1}^K \!\alpha_k e^{j \!\frac{4\pi\! f_0\!v_k}{c}\!nT_s}\! e^{\!-\!j\!\frac{4\pi \!f_m\!r_k}{c}}
\!\mathbf{w}_{\!R\!X\!,q}^H \mathbf{a}_{R\!X}(\!\theta_k\!)\mathbf{a}_{T\!X}^H(\!\theta_k\!)\tilde{\mathbf{x}}_{q\!,n\!,m} \!\!\right] \\ & \quad \quad +
\sum_{i=1}^I  \beta_i e^{-j2\pi f_m\!\frac{2r_i'}{c}} \mathbf{w}_{RX,q}^H \mathbf{a}_{R\!X}(\theta_i')\mathbf{a}_{T\!X}^H(\theta_i') \tilde{\mathbf{x}}_{q,n,m}
\\&= \!\sum_{k=1}^{K}\!\left[ \! \frac{1}{N}\!\! \sum_{n=0}^{N-1}\!\! \alpha_k e^{j \!\frac{4\pi\! f_0\!v_k}{c}\!nT_s}\! e^{\!-\!j\!\frac{4\pi \!f_m\!r_k}{c}}
\!\mathbf{w}_{\!R\!X\!,q}^H \mathbf{a}_{R\!X}(\!\theta_k\!)\mathbf{a}_{T\!X}^H(\!\theta_k\!)\tilde{\mathbf{x}}_{q\!,n\!,m} \right]  \\& \quad \quad + \sum_{i=1}^I  \beta_i e^{-j2\pi f_m\!\frac{2r_i'}{c}} \mathbf{w}_{RX,q}^H \mathbf{a}_{R\!X}(\theta_i')\mathbf{a}_{T\!X}^	H(\theta_i') \tilde{\mathbf{x}}_{q,n,m}
\\&=\!\sum_{k=1}^{K}\!\!\left[\!\alpha_k 
e^{\!-\!j\!\frac{4\pi \!f_m\!r_k}{c}}
\!\mathbf{w}_{\!R\!X\!,q}^H \mathbf{a}_{R\!X}(\!\theta_k\!)\mathbf{a}_{T\!X}^H(\!\theta_k\!)\tilde{\mathbf{x}}_{q\!,n\!,m}
\!\frac{1}{N}\!\!\!\sum_{n=0}^{N-1} \!\!\! e^{\!j\!\frac{4\pi\! f_0\!v_k}{c}\!nT_s}\!\right] \\& \quad \quad + \sum_{i=1}^I  \beta_i e^{-j2\pi f_m\!\frac{2r_i'}{c}} \mathbf{w}_{RX,q}^H \mathbf{a}_{R\!X}(\theta_i')\mathbf{a}_{T\!X}^	H(\theta_i') \tilde{\mathbf{x}}_{q,n,m}.
		\end{aligned}
	\end{split}\tag{47}
\end{equation}
\textcolor{black}{Since}
\begin{equation}
	\begin{split}\color{black}
		\begin{aligned}
			\label{deqn_ex1a}
&\lim\limits_{N \to \infty}\frac{1}{N}\sum_{n=0}^{N-1} e^{j\frac{4\pi f_0v_k}{c}nT_s} \\&   =  \lim\limits_{N \to \infty} \left[\frac{1}{N}
e^{j\frac{2\pi f_0v_kT_s(N-1)}{c}}
\frac{\sin (2\pi f_0v_kT_sN/c)}{\sin (2\pi f_0v_kT_s/c)} \right]=0,
		\end{aligned}
	\end{split}\tag{48}
\end{equation}
\textcolor{black}{there is} 
\begin{equation}
	\begin{split}\color{black}
		\begin{aligned}
			\label{deqn_ex1a}
&\lim\limits_{N \to \infty} \tilde{h}^{static}_{m,q} = \lim\limits_{N \to \infty} \frac{1}{N}\sum_{n=0}^{N-1} \tilde{h}^{s,q}_{n,m} \\& =  \sum_{i=1}^I  \beta_i e^{-j2\pi f_m\!\frac{2r_i'}{c}} \mathbf{w}_{RX,q}^H \mathbf{a}_{R\!X}(\theta_i')\mathbf{a}_{T\!X}^	H(\theta_i') \tilde{\mathbf{x}}_{q,n,m}
\\& = \mathbf{w}_{RX,q}^H
\left(\sum_{i=1}^I \mathbf{H}_{i,n,m}'\right)\tilde{\mathbf{x}}_{q,n,m}.
		\end{aligned}
	\end{split}\tag{49}
\end{equation}
\textcolor{black}{Then Eq. (36) has been proven.}

\bibliographystyle{ieeetr}
\bibliography{paper3.bib}

\vfill

\end{document}